\documentclass[twoside]{elsart}
\journal{Astroparticle Physics}
\usepackage{hangcaption}
\usepackage[dvips]{graphicx}
\usepackage[dvips]{color}
\usepackage{natbib}
\usepackage{amssymb}  
\begin{document}
\def\dl{\Phi^{\mbox{DL}}}
\def\stl{\Phi^{\mbox{SL}}}
\def\sdl{\Phi^{\mbox{SDL}}}
\def\ds{\Phi^{\mbox{DS}}}
\def\ps{\Phi^{\mbox{PS}}}
\def\diffunits{\mbox{GeV cm}^{-2}\,\mbox{s}^{-1}\,\mbox{sr}^{-1}}
\def\pointunits{\mbox{GeV cm}^{-2}\,\mbox{s}^{-1}}
\def\en{E_{\nu}}
\def\eg{E_{\gamma}}
\def\ep{E_{p}}
\def\epb{\epsilon_{p}^{b}}
\def\enb{\epsilon_{\nu}^{b}}
\def\enbG{\epsilon_{\nu,GeV}^{b}}
\def\enbM{\epsilon_{\nu,MeV}^{b}}
\def\ens{\epsilon_{\nu}^{s}}
\def\ensG{\epsilon_{\nu,GeV}^{s}}
\def\egb{\epsilon_{\gamma}^{b}}
\def\egbM{\epsilon_{\gamma,MeV}^{b}}
\def\g25{\Gamma_{2.5}}
\def\lumi{L_{\gamma}^{52}}
\def\dunits{\mbox{GeV}\,\mbox{cm}^{-2}\,\mbox{s}^{-1}\,\mbox{sr}^{-1}}
\begin{frontmatter}
\title{Astrophysical implications of high energy neutrino limits}
\author[dort]{Julia K.~Becker\corauthref{cor}}
\author[dort,heid]{, Andreas Gro{\ss}}
\author[dort]{, Kirsten M\"unich}
\author[dort]{Jens Dreyer}
\author[dort]{, Wolfgang Rhode}
\author[bonn,bonn2,alabama]{, Peter L.~Biermann}
\corauth[cor]{{\scriptsize Corresponding author. Contact: julia.becker@udo.edu, phone: +49-231-7553667}}
\address[dort]{Universit\"at Dortmund, Institut f\"ur Physik, D-44221
  Dortmund, Germany}
\address[heid]{Max-Planck-Institut f\"ur Kernphysik, D-69177 Heidelberg, Germany}
\address[bonn]{Max Planck Institut f\"ur Radioastronomie, Auf dem H\"ugel 69,
  D-53121 Bonn, Germany}
\address[bonn2]{Department of Physics and Astronomy, University of Bonn,
  Germany}
\address[alabama]{Department of Physics and Astronomy, University of Alabama,
  Tuscaloosa, Alabama 35487, USA}
\date{\today}

\begin{abstract}
Second generation high energy neutrino telescopes are being built to
reach sensitivities of neutrino emission from galactic and extragalactic
sources. Current neutrino detectors are already able to set limits which are 
in the range of some emission models. In particular, the {\bf A}ntarctic {\bf
  M}uon {\bf A}nd {\bf N}eutrino {\bf D}etection {\bf A}rray ({\sc AMANDA})
has recently presented the so far most restrictive limit on diffuse neutrino
emission~\citep{diffuse_4yrs}. Stacking limits which apply to AGN point source
classes rather than to single point sources~\citeauthor{andreas} are given as well. In this paper, the
two different types of limits will be used to draw conclusions about different
emission models. An interpretation of stacking limits as diffuse
limits to the emission from considered point source class is presented. The limits
can for instance be used to constrain the predicted correlation of
{\sc EGRET}-detected diffuse emission and neutrino emission. Also, the
correlation between X-ray and neutrino emission is constrained. Further
results for source classes like
TeV blazars and FR-II galaxies are presented. Starting from the source catalogs so-far examined for the stacking
method, we discuss further potential catalogs and examine the possibilities of
the second generation telescopes {\sc IceCube} and {\sc KM3NeT} by comparing catalogs with
respect to northern and southern hemisphere total flux.
\end{abstract}
\end{frontmatter}

\section{Introduction \label{introduction}}
As is well known for hundreds of years, the sky would be infinitely bright
in optical light, if all space were homogeneously full of stars, and space
were infinitely extended without any change in properties~\citep{olbers}.  This reasoning
can be applied to stars as well as to galaxies. This argument can also be
applied to electromagnetic emissions and just as well to neutrino emission.  The
solution to Olbers paradox is that space is not homogeneously full of identical
distributions of stars or galaxies, and that the Universe evolves quite
strongly.  This implies that the integral over the optical light of
all galaxies has a finite sum, which does not exceed an observable
level.  Clearly, given a proper sampling of galaxies, it is
possible to determine this level. In some cases, there is a unique
relationship between the optical light and the emission at some other
wavelength. Starting from this correlation, the summed emission from all
galaxies over all history can be deduced and it is a simple function of the sum at 
optical wavelengths. The optical emission can certainly be replaced with the
emission at some other electromagnetic wavelength, such as X-rays. Equally,
the correlated wavelength can be neutrino emission. If we
now do not know what the sum of the emission is at the electromagnetic
wavelength, but have limits, then a limit 
for the neutrino emission can be deduced a fortiori. Vice versa, if we have a limit for the
neutrino emission, a corresponding limit for the
electromagnetic emission can be derived.  In either case, it might happen, that we have a
known background, and then can deduce whether this specific class of
sources could possibly explain all background at the other
"wavelength". We start this argument with neutrinos, set a limit, and
then derive a limit for the electromagnetic background. We could also
start with the electromagnetic background, and then set a limit for the
neutrino emission. So we have an {\em Olbers paradox for neutrinos}.  This is
the key point of this paper.

More concretely, the interpretation of different {\sc AMANDA} neutrino flux limits is done for
different classes of Active Galactic Nuclei (AGN). In the case of a stacking
analysis in which the strongest sources from the same class are selected, a
method for the interpretation of these limits as limits to the diffuse flux
from the given source class is developed. 
Additionally, future possibilities of second generation neutrino telescopes
like {\sc IceCube} and {\sc KM3NeT} are examined. Different source classes are
discussed with respect to their distribution in the sky, i.e.~which sources
are in the northern and which are in the southern hemisphere. The field of
view for {\sc IceCube} respectively {\sc KM3NeT} is the northern respectively
the southern hemisphere and their view of the sky is complementary.

This paper is organized as follows: in Section~\ref{nu_models}, the
prevailing models are discussed according to the
normalization options given by experimental observations of Cosmic Rays (CRs)
and photons at different wavelengths. It is essential to discuss in this
context which stacking limit can be applied to what prediction. Current neutrino flux
limits are investigated in Section~\ref{nu_limits:sec}. A general ansatz for the interpretation of 
stacking limits as diffuse limits is presented here. In Section~\ref{stack_results},
the method is applied to {\sc AMANDA}'s limits which have been derived by stacking
AGN according to their electromagnetic output. 
The question which stacking limits apply to which normalization scenario is
discussed in Section~\ref{compare} and it
is examined how the limits constrain different
models.
In Section~\ref{class_analysis}, further source classes are examined
according to the possibility of applying the stacking method for high energy
neutrinos, i.e.~$\en>$~TeV. In addition, a comparison of the contribution from the northern and
southern hemisphere is done for various source classes in order to compare the
capabilities of next generation's telescopes {\sc IceCube} and
{\sc KM3NeT}. The results are summarized in Section~\ref{conclusions}.

\section{Discussion of prevailing neutrino flux models \label{nu_models}}
In some hadronic acceleration models
it is assumed that for each class of AGN, the electromagnetic emission is
correlated to a neutrino signal. Apart from individual normalization factors, the
corresponding cosmological integrations are basically mathematically
identical. In this section, the correlation between the emission of neutrinos
and photons at different wavelengths will be discussed according to neutrino
flux models which are currently being discussed in the literature.
\subsection{TeV photon sources and Cosmic Rays}
Sources of electromagnetic TeV emission can
be interpreted as optically thin to photon-neutron interactions,
$\tau_{\gamma\,n}\ll 1$ \citep[e.g.][]{muecke,mpr} in hadronic acceleration models. In such a
scenario, charged Cosmic Rays (CRs) are produced in the vicinity of the source
through the decay of the escaping neutrons. In this case, the resulting
neutrino energy density would be proportional to the extragalactic CR component measured
at Earth. A theoretical upper bound of such a contribution to the diffuse
neutrino flux has been derived by Mannheim, Protheroe and Rachen (in the
following referred to as the {\em MPR bound}), see \citep{mpr}. Within the framework of the proton-blazar model, the neutrino flux from High-frequency peaked BL Lacs (HBLs) has been calculated
using the connection between Cosmic Rays and neutrino emission,
see~\citep{muecke}.
\subsection{Normalization to the diffuse {\sc EGRET} and {\sc COMPTEL} spectrum}
In the case of optically thick sources ($\tau_{\gamma\,n}>1$), the photons interact with nucleons in
the source before escaping at lower energies leading to the emission of
sub-TeV photons. Therefore, the diffuse
extragalactic background determined by the {\sc EGRET}\footnote{{\bf E}nergetic {\bf
    G}amma {\bf R}ay {\bf E}xperiment {\bf T}elescope} 
experiment~\citep{sreekumar98,egret_new} ($\eg>100$~MeV) can be interpreted as an avalanched TeV
signal from blazars and can thus be used to normalize the neutrino flux from
{\sc EGRET} or {\sc COMPTEL}\footnote{{\bf COM}pton {\bf TEL}escope}-type sources. Again, MPR give an upper bound to the contribution from such
sources in~\citep{mpr}, which is much less restrictive than the optically thin
case. Apart from the bound, a calculation of the maximum contribution from
blazars is given in~\citep{mpr}. 
In addition to the contribution from HBLs within the framework of the proton-blazar
model, a contribution from the optically thick Low-frequency peaked BL Lacs (LBLs)
can be calculated using the {\sc EGRET} diffuse extragalactic background for a
normalization of the neutrino spectrum, see~\citep{muecke}.
A model of $p\,\gamma$ interactions in AGN and
collisions of protons from the core with protons of the host galaxy is derived
in~\citep{mannheimjet}. This model will be referred to as {\em
  M95-A}. Alternatively to optically thin sources with photon emission above
$100$~MeV, the environment can be optically thick allowing only
photons below $100$~MeV to escape. In that case, the neutrino signal can be
normalized to the diffuse extragalactic contribution measured by {\sc COMPTEL} at
energies in the range of $(0.8,30)$~MeV~\citep{comptel}. This would enhance the
contribution of neutrinos from $p\,\gamma$ by almost an order of magnitude as
shown in~\citep{mannheimjet}. We will refer to this model as {\em M95-B} in
the following. 

A model by~\cite{stecker96} was originally using the diffuse cosmic X-ray background (see
Sec.~\ref{rosat_prediction}) and has recently been modified in a way that it is using the
{\sc COMPTEL} diffuse background to normalize the neutrino spectrum~\citep{stecker_mod}. This reduces
the formerly very high contribution by a factor of 10. In addition, oscillations have been taken
into account which leads to a further reduction by a factor of 2. 
\subsection{The {\sc ROSAT} X-ray background as normalization option
  \label{rosat_prediction}}
The measurement of the diffuse extragalactic contribution in X-rays by {\sc
  RO\-SAT}\footnote{{\bf RO}entgen {\bf SAT}ellite}
has raised the question whether it is produced by radio-weak AGN. Assuming that the
X-ray emission comes from the foot of the jet, the X-ray
signal would be accompanied by a neutrino flux. A model by~\cite{nellen} and an approach by~\cite{stecker96} have
been presented. An alternative scenario would be the up-scattering of thermal
electrons via the Inverse Compton effect. In that case, the X-ray component
would not be accompanied by a neutrino signal, see e.g~\citep{msr}. Until today, about $75\%$ of the diffuse
X-ray signal has been resolved by {\sc ROSAT}~\citep{xray_sources}, with the
help of {\sc Chandra} and {\sc XMM\footnote{{\bf X}-ray {\bf M}ulti-{\bf M}irror Mission}-Newton} data,
this number can be updated to $90\%$, see~\citep{xray_sources} and references therein.  More than $70\%$ of the diffuse background
could be connected to the X-ray emission of Active Galactic Nuclei most of
which are radio weak. {\sc AMANDA}'s measurements of a diffuse neutrino flux
did not yield a significant signal and constrains these models strongly. 
\subsection{Correlated radio and neutrino emission}
Detailed examination of multi-wavelength observations have shown a
 correlation between the disk and jet power of Active Galactic Nuclei,
see~\citep{falcke,falcke1,falcke3}. This correlation can be used to determine the
diffuse neutrino flux from FR-II galaxies and flat spectrum radio quasars within the framework of
the jet-disk symbiosis model~\citep{bbr_05}. The neutrino
flux has been assumed to follow an $E^{-2}$ spectrum which leads to a
relatively high neutrino flux at energies above TeV energies. Assuming on the other hand a correlation between
the radio spectral index and the index of the proton spectrum leads to an
$E^{-2.6}$ neutrino spectrum. Since the flux is normalized at low energies,
the contribution decreases significantly at higher energies with an increasing
spectral index. 
\section{Neutrino flux limits\label{nu_limits:sec}}
The conversion of stacking limits into stacking diffuse limits is the main
topic of this section. Basically we use the contribution of the identified
sources to the integrated background at a chosen electromagnetic wavelength to
estimate by way of a physical model the corresponding ratio for neutrinos:
What fraction of identified neutrino source candidates goes towards the
integrated background? It is important to note here that we use conservative estimates in order to obtain absolute, reliable upper limits. As 
limits to the flux of identified source classes we use the sample of AGN
classes from the {\sc AMANDA} stacking analysis which is described in~\citep{andreas,5yrs}.

To avoid any confusion of the notation, a paragraph on
the limit conventions precedes the actual discussion of the limits.
\subsection{Limit Conventions}
Throughout this paper, several types of neutrino flux limits
  will appear. Limits are usually given in the form $E^2\cdot dN/dE$
  with $dN/dE$ as the differential neutrino flux at Earth and E as the
  neutrino energy. 

\begin{figure}[h!]
\centering{
\includegraphics[width=\linewidth]{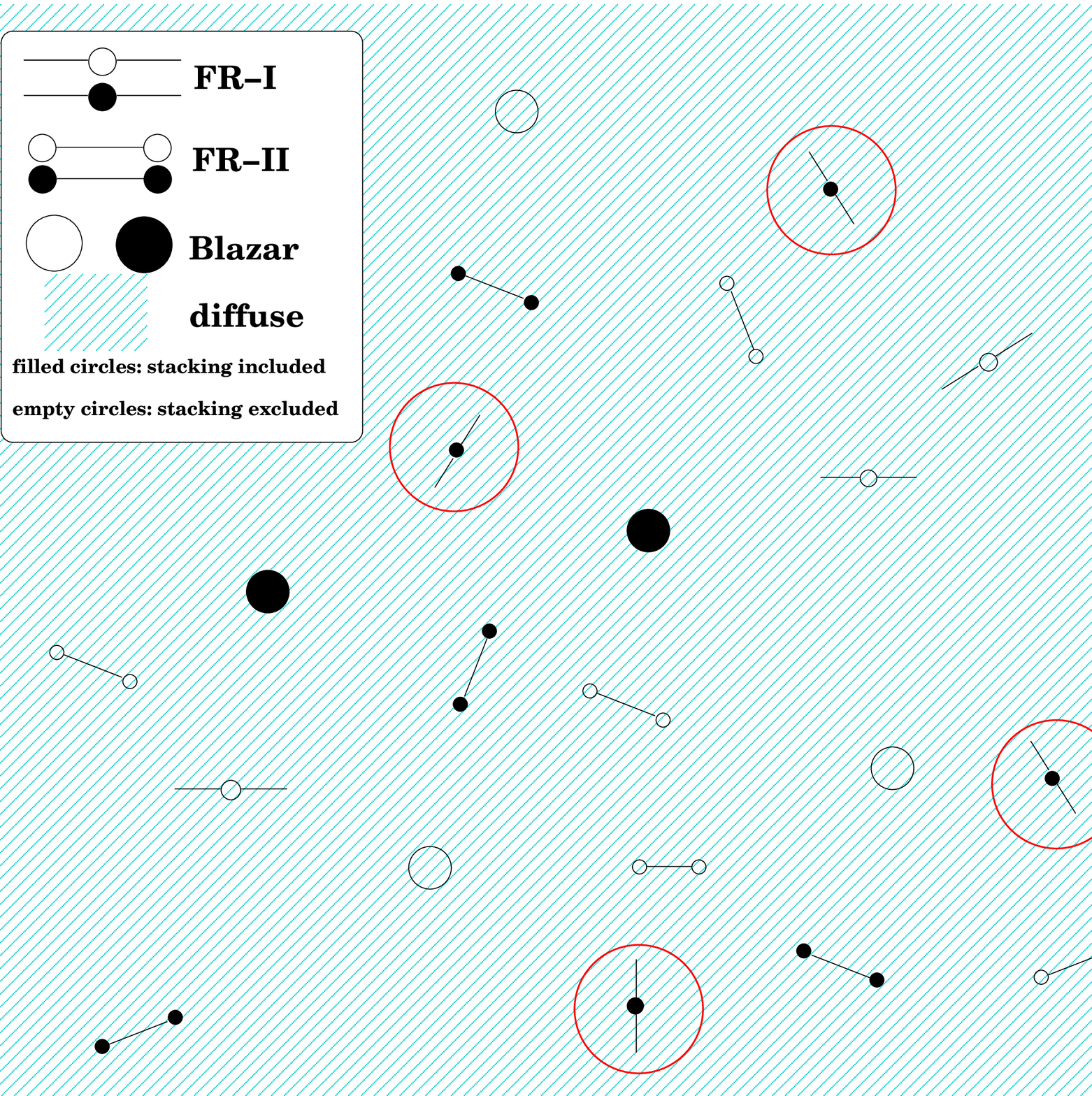}
\hangcaption{Schematic figure of the principle of source stacking. Objects are
classified into source classes (here: FR-I and FR-II galaxies as well as
blazars). Under the signal hypothesis that photons in a certain bandpass are
directly correlated to neutrinos, the sources are selected according to their
strength in the corresponding bandpass. The most intense sources are then
stacked for an analysis of a neutrino signal from the source class. This is
indicated for the case of FR-I galaxies as circles around the objects. Only
the strongest sources are used in the stacking analysis (filled circles),
while weaker objects are not used (empty circles). An estimate of the
remaining source signal is needed as well as knowledge about unresolved
sources, possibly seen as diffuse background in order to interpret the
stacking neutrino limits as diffuse upper bounds to the given source
class. The diffuse background varies with the source class.\label{stacking_diffuse_fig}}
}
\end{figure}

Throughout the paper, the limits will be denoted as follows:
\begin{itemize}
\item $\dl$: Diffuse Limit (DL) derived by using data
  from the complete northern hemisphere. It is given in units of $\diffunits$. 
\item $\stl$: Stacking Limit (SL) in units of $\pointunits$, obtained for the point source flux from a
  certain class of AGN. The principle is indicated schematically in
  Fig.~\ref{stacking_diffuse_fig}. As an example, three source classes, FR-I
  and FR-II galaxies as well as blazars, are displayed. The stacking limits
  are obtained by stacking the most luminous sources of the same class in the
  sky, indicated by sources with filled circles. Weaker sources (empty
  circles) are not included. 
\item $\sdl$: Stacking Diffuse Limit (SDL), derived from the stacking limit in
  the same units as the diffuse limit, $\diffunits$. It is determined by
  taking into account the contribution from weaker sources as well as yet
  unidentified sources, present in a diffuse background.
\end{itemize}
A similar convention is used to denote the corresponding sensitivities:
\begin{itemize}
\item $\ds$: Diffuse Sensitivity (DS), giving the sensitivity of
  {\sc AMANDA} towards a diffuse signal from the northern hemisphere. Units are $\diffunits$. 
\item $\ps$: the sensitivity to a single point source in {\sc AMANDA}. Units are $\pointunits$.
\end{itemize}
\subsection{Limits on the total diffuse neutrino flux}
The unfolded spectrum of the diffuse neutrino flux
as observed by the \\{\sc AMANDA} experiment is indicated in e.g.~Fig.~\ref{blazars_thin}. It can be seen that the measured flux follows the prediction
of the atmospheric neutrino spectrum and no extragalactic contribution is
observed \citep{kirsten_icrc05}. The unfolded spectrum is derived using data
from the year 2000. The current limit on the diffuse muon neutrino flux from
extragalactic sources is given for four years of data (2000-2004)~\citep{diffuse_4yrs},
\begin{equation}
\dl=7.4\cdot 10^{-8}\dunits
\end{equation}
in the energy range of $10^{4.2}$~GeV to $10^{6.4}$~GeV.

In addition to a diffuse search, {\sc AMANDA} examines a possible signal from single
sources. No signal excess above the atmospheric background was
observed yet. To maximize the significance of the signal to background expectation, a stacking method has been applied
examining different source classes of AGN~\citep{andreas}. This method is commonly used in cases where a single source is not likely to 
have a significant signal above background \citep[e.g.][]{beacons}. Given the limit for a certain source class, $\stl$, the differential flux renormalized to
a diffuse signal is 
\begin{equation}
\sdl= \epsilon \cdot \xi \cdot \frac{\stl}{2\pi\mbox{sr}}\,.
\end{equation}
The  transition from point source to diffuse flux is done by dividing the source
limits by {\sc AMANDA}'s field of view, $2\,\pi$~sr. 
The stacking factor $\epsilon$ is the ratio of the total
photon signal in the sample and the photon signal which is included in the
analysis. Figure~\ref{stacking_diffuse_fig} gives a schematic
representation. The stacking factor corrects for all sources with empty circles in
the diffuse limit, since the signal from these sources is not included in the
stacking analysis. 
The diffusive factor $\xi$ represents the ratio of the total diffuse flux
to the contribution of resolved sources. Sources
which have not been resolved yet have to be considered as potentially
contributing to the total diffuse neutrino signal as they contribute to the
isotropic photon signal. In Fig.~\ref{stacking_diffuse_fig}, this is indicated
by the diffuse background which is different for each source class. This
background is identified in a few cases by looking at the luminosity function of the
objects, in others by measurements of the diffuse component. A detailed view
on $\epsilon$ and $\xi$ will be given in Section~\ref{stack_results}. 
\section{Point source limits on AGN neutrino fluxes \label{stack_results}}
\begin{figure}
\centering{
\includegraphics[width=10cm]{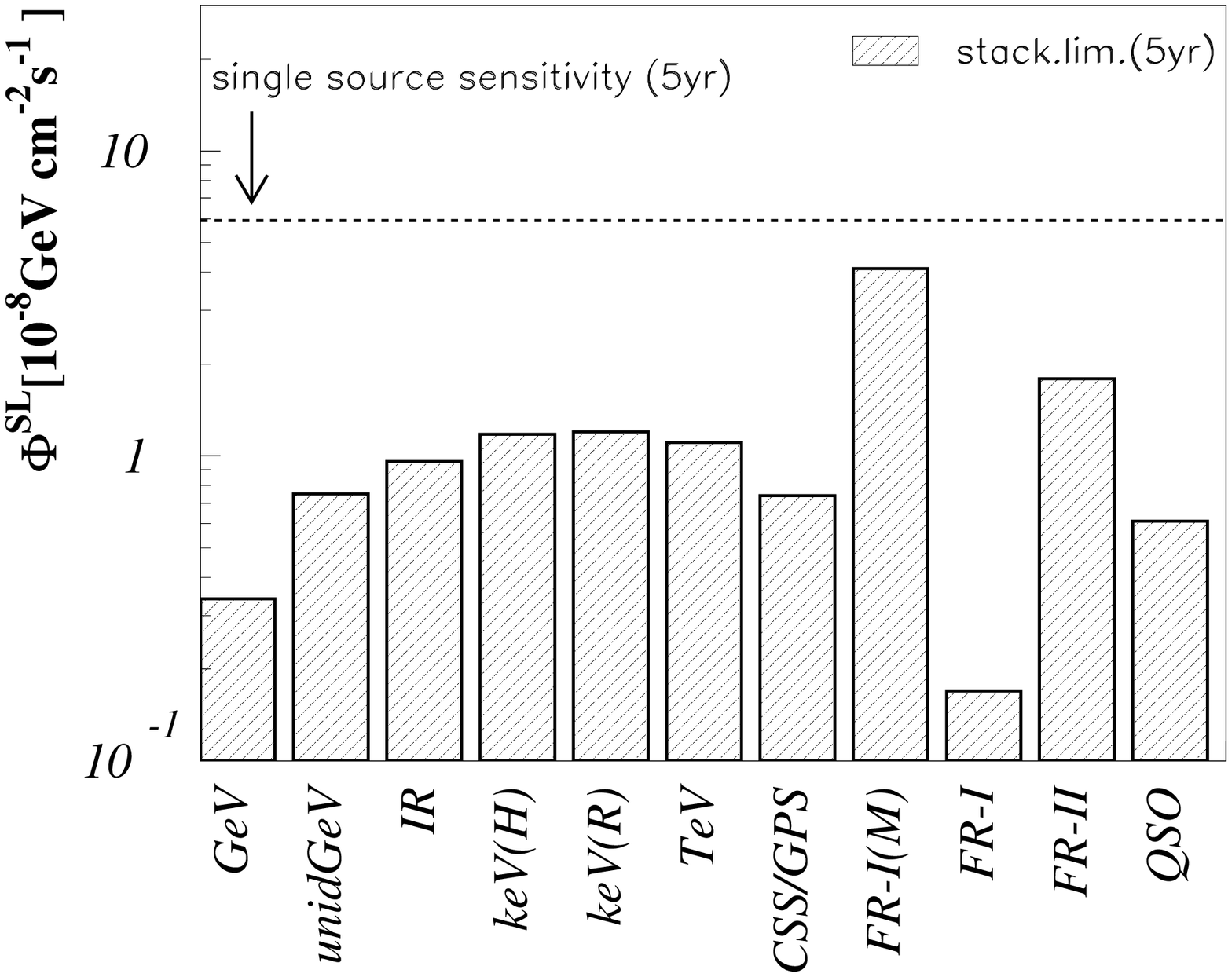}
\hangcaption{Limits on the neutrino flux for a given source class. Limits in this representation are
  given per source. The source classes are labeled with the abbreviation
  indicated in the list of the samples above.}
\label{5yr_stack_fig}
}
\end{figure}

Recently, a stacking analysis has been published for the first time for a neutrino signal from Active
Galactic Nuclei for which eleven AGN source classes have been defined. The source
classes have been defined and selected as follows. The catalogs below have
been selected in~\citep{andreas}.
\begin{enumerate}
\item GeV blazars detected by {\sc EGRET} ({\em GeV}).
\item Unidentified GeV sources observed by {\sc EGRET} ({\em
    unidGeV}). Sources of possible galactic origin have been excluded.
\item Blazars that have been observed at infrared
  wavelengths ({\em IR}).
\item Blazars that have been observed at keV wavelengths by the experiment
  {\sc HEAO-A}\footnote{{\bf H}igh {\bf E}nergy {\bf A}stronomy
  {\bf O}bservatories{\bf -A}} ({\em keV(H)}).
\item keV blazars observed by the {\sc ROSAT} experiment ({\em keV (R)}).
\item The class of blazars with observed TeV emission ({\em TeV}).
\item Compact Steep Sources (CSS) and GHz peaked sources (GPS) ({\em CSS/GPS})
  as selected by~\cite{odea}.
\item Low luminosity Faranoff Riley galaxies (FR-I) with M-87
  included ({\em FR-I(M)}). 
\item FR-I galaxies excluding M-87 ({\em FR-I}). These two different cases are necessary to
  consider, since M-87 is the closest AGN (distance $\sim 20$~Mpc) and dominates the total signal of
  all FR-I galaxies.
\item High luminosity Faranoff Riley galaxies ({\em FR-II}).
\item Quasi Stellar Objects ({\em QSO}).
\end{enumerate}
 References to the
corresponding catalogs as well as a detailed description of the analysis
methods are given there. In brackets, the
abbreviations for the source classes are given as they appear in
Fig.~\ref{5yr_stack_fig}. 

The limits on the neutrino flux from the given source classes are displayed in
Fig.~\ref{5yr_stack_fig}. They are compared to {\sc AMANDA}'s single source
sensitivity of $\ps=5.9\cdot 10^{-8}\,\pointunits$
to point out the improvement that has been achieved in the stacking approach. In
the following, the stacking factor $\epsilon$ and the diffusive factor $\xi$
will be discussed. It
will be shown that seven of the eleven samples can effectively be used to
determine a diffuse limit on the neutrino flux from the given class. 
\subsection{The Stacking factor}
The number of sources in a sample that is included in the stacking analysis is
determined by optimizing the expected significance for signal and background. Subsequently, the amount of
the total photon signal of the analyzed sources is determined by the number of included
objects. This is shown schematically in Fig.~\ref{stacking_diffuse_fig}. The
strongest FR-I
galaxies (filled circles) are included in the stacking analysis. The stacking
process for FR-I galaxies is indicated by circles around the stacked objects. FR-I
galaxies having a too weak signal (empty circles) are not part of the stacking
and have thus to be included in the {\em stacking factor}
$\epsilon$. Quantitatively, the ratio of
the total photon signal and the signal of the sources that are used in the
stacking approach is represented by $\epsilon$. The
precise values of $\epsilon$ for
each class are given in Table~\ref{5yr_stack_tab}. 

Some of the stacking classes cannot
be used for a diffuse interpretation. Firstly, the
number of sources needs to be sufficiently high. An indication for an
insufficient number of sources would be the use of all the sources in a sample in the
stacking analysis. Secondly, the luminosity
function should be moderately falling: if only very few sources in our vicinity
dominate the signal, no conclusions about the overall diffuse contribution can
be justified.

The complete sample was used for stacking in several cases due to the low number of detected sources in the
given wavelength interval. In particular,
these source classes are the infrared and {\sc HEAO-A} sources as well as the TeV
blazars. Since all twelve observed infrared sources have been used and only three sources
are reported from {\sc HEAO-A}, a diffuse limit
to the source class cannot be derived for these samples. There are five sources in the stacking sample of TeV
blazars. Due to the small number, this class cannot be used for a diffuse
interpretation. Due to the strong absorption of the TeV signal in photons, the
contribution of a neutrino flux from TeV-resolved photon sources is expected
to be much lower than the total flux. This is discussed in
detail in section~\ref{compare}.

Apart from the three classes already excluded, the class of FR-I galaxies
including M-87 is not suitable for an interpretation of a stacking limit as a
diffuse limit: M-87 makes up most of the photon signal from all FR-I
galaxies. This is also the reason why the stacking limit is very close to the
single point source sensitivity. This result must therefore rather be
interpreted as a point source result of M-87. The FR-I sample without M-87 on
the other hand is applicable for the diffuse interpretation.

\begin{table}[h!]
\begin{tabular}{l|lllll}
Source class&selected&$\epsilon$&$\xi$&$\stl$                         &possible\\
            &wavelength&        &     &&diffuse
            limit?\\\hline\hline
{\sc EGRET} GeV&$>100$~MeV&1.4&$<12$&2.71&yes\\
unid. {\sc EGRET}&$>100$~MeV&1.3&1&31.7&yes\\
infrared&$60\,\mu$m &1&-&10.6&no\\
keV ({\sc HEAO-A})&$(0.25,25)$~keV&1&-&3.55&no\\
keV ({\sc ROSAT})&$(0.2,2)$~keV&1.2&1.43&9.71&yes\\
TeV blazars&$>100$~GeV&1&-&5.53&no\\
CSS/GPS&178~MHz,&1.3&model dep.&5.94&yes\\
       &2.7~GHz,5~GHz& & & & \\
FR-I w.~M-87&178~MHz&-&-&4.11&no\\
FR-I w/o~M-87&178~MHz&1.1&model dep.&2.91&yes\\
FR-II&178~MHz&2.65&$<160$&30.4&yes\\
QSOs&UV&1.3&model dep.&6.70&yes\\
\end{tabular}
\caption{Table of the source class limits obtained with the stacking 
method. Five years of data, 2000-2004, have been used for the analysis with
{\sc AMANDA} in \citep{5yrs}. Listed are the source class, the selection
wavelength, the stacking and diffusive factors, the stacking limit $\stl$ in
units of $10^{-8}\,\pointunits$ and the possibility of interpreting the
stacking result as a diffuse limit.}
\label{5yr_stack_tab}
\end{table}
\subsection{The diffusive factor}
Apart from the contribution of resolved sources to a diffuse background, a
component of unresolved sources has to be considered. The ratio of the
total diffuse signal to the signal of resolved sources is called the {\em diffusive
factor} $\xi$ in the
following. In Fig.~\ref{stacking_diffuse_fig}, this is indicated as the
diffuse photon background of unidentified sources. A truly diffuse contribution from e.g.~dark matter decays is discussed as a
component of the diffuse photon background~\citep{bk06}, but it will be considered as
negligible here. Any contribution of a diffuse component not connected to any
source population would improve the limits accomplished.
  
$70\%$ of the diffuse X-ray background has
been identified as AGN in the {\sc ROSAT} data sample~\citep{xray_sources}. Thus, the diffusive factor $\xi$,
which is the inverse of the fraction of resolved sources, for the keV {\sc ROSAT}
data sample is given as $\xi_{ROSAT}=1.43$. Note that today, {\sc XMM-Newton}
and {\sc Chandra} provide values of $\sim 95\%$ resolved sources. But since both the
neutrino flux predictions are based on and the
stacking limit has been derived from {\sc ROSAT} data, a fraction of $70\%$
resolved sources is used.

For {\sc EGRET} sources, it could be shown that only 1/12 of the total diffuse
background can be made up by resolved sources, as derived by~\cite{chiang}. The claim from~\cite{chiang} that about $25\%$ of the 
diffuse background of blazars is made up by resolved AGN, while the remaining contribution is
 from unresolved sources which are not blazars, would reduce $\xi$
significantly to $\xi_{EGRET}\approx 4$. However, \cite{stecker_salamon01} point out that this estimate is based on
assumptions that do not hold for {\sc EGRET} blazars. Thus,
$\xi_{\max}=12$ will used as a conservative number in the following. With the
launch of {\sc GLAST}\footnote{{\bf G}amma-ray {\bf
    L}arge {\bf A}rea {\bf S}pace {\bf T}elescope}, the question of the contribution from resolved blazars
will most likely be settled.


For the remaining source classes, there is no explicit estimate of the
contribution of resolved sources to a diffuse background. In this case, an
estimate of $\xi$ can be possible for source populations where there is an
estimate of the weakest sources in the sky. In this case, the luminosity
evolution of the population has to be considered. In the Euclidean case,
sources evolve as $N(>S)\propto S^{-3/2}$ with $S$ as the observed flux and
$N(>S)$ as the number of sources for fluxes greater than $S$. In such a case, the integral of the
distribution diverges for a vanishing minimal source strength,
$N(S_{\min}\rightarrow 0)\rightarrow 0$. For cosmological sources, the
behavior is not Euclidean and in some cases, it is possible to determine a
lower threshold for the flux, see \citep[e.g.]{longair_galaxy_formation}. For
instance, such determination can be done in the case of
FR-II galaxies which are high luminous AGN with a defined luminosity limit of
$L_{radio}\sim 10^{43}$~erg/s~\citep{fr}.

\begin{figure}[h!]
\centering{
\includegraphics[width=12cm]{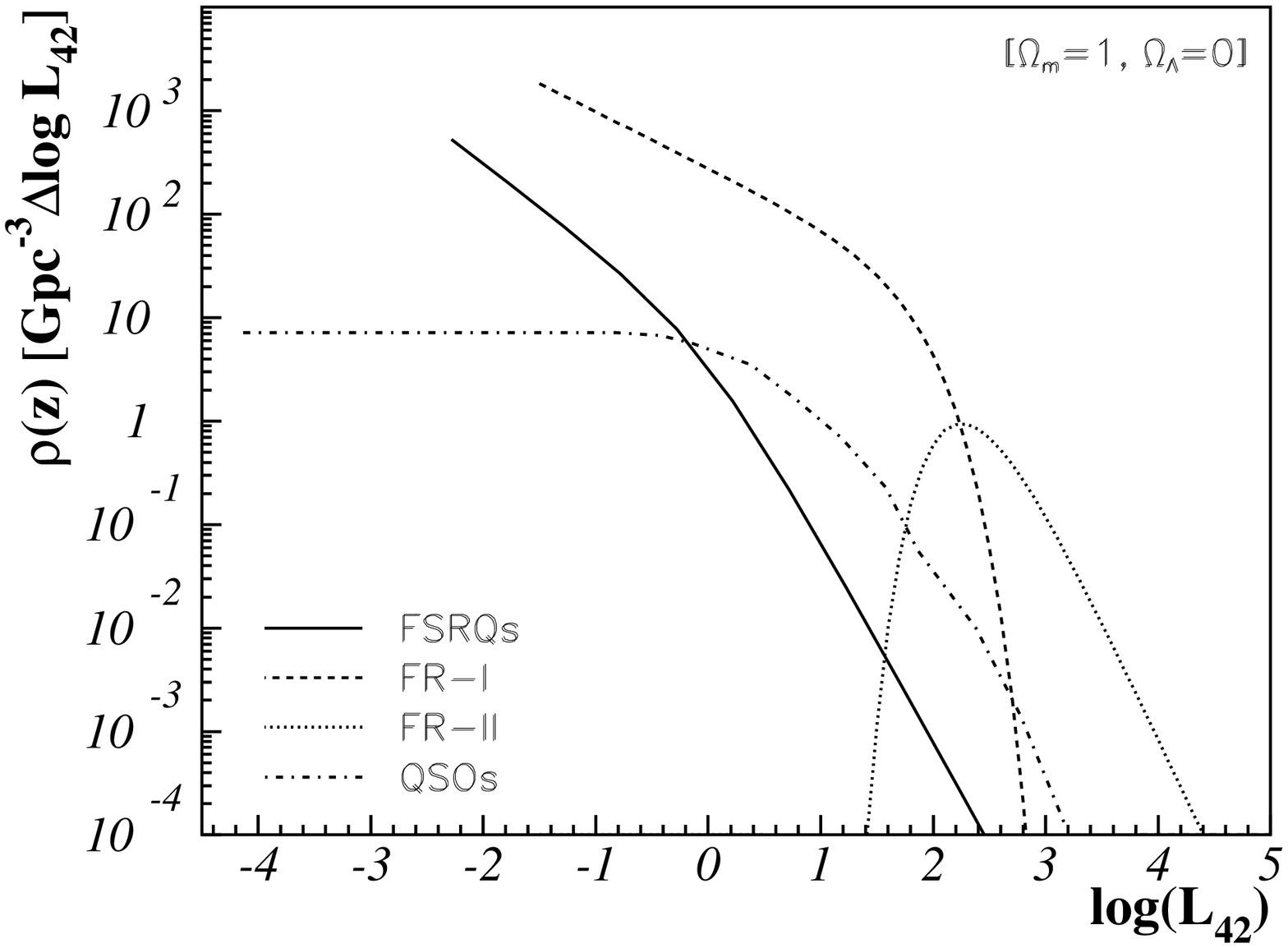}
\hangcaption{Radio Luminosity Functions at $z=0$ for Flat Spectrum Radio Quasars
  (FSRQs, solid line)~\citep{dunlop}, FR-I and FR-II galaxies (dashed and
  dotted lines)~\citep{willott}, as well as for
  QSOs (dot-dashed line)~\cite{schmidt72}. Density units are 1/Gpc$^3\cdot
  \Delta\log L$ and luminosity units are
  $L_{42}:=L/(10^{42}$~erg/s). Only FR-II galaxies show a cutoff
  at low luminosities. All other source classes show an increasing or constant
  population with decreasing luminosity.}
\label{lumi_fcts}
}
\end{figure}

The determination of a lower luminosity limit enables the calculation of the
total number of sources in a population $N_{tot}$. The total number of sources
is given by integrating the number per flux interval, dN/dS over the flux:
\begin{equation}
N_{tot}=\int_{S_{\min}}^{\infty}\frac{dN}{dS}\,dS\,.
\end{equation}
For the case of divergence for $S_{\min}\rightarrow 0$, the lower integration
limit is essential to know as mentioned before. This is not trivial, since
measurements for most source classes do not reveal the low luminosity
cutoff. The luminosity functions of different AGN classes at $z=0$ are shown
in Fig.~\ref{lumi_fcts}.

In this case, the diffusive factor
can be estimated by assuming that sources which are not in the catalog are not
stronger than the weakest source in the catalog. Thus, the flux of any
additional source $S_{add}^{i}$ is smaller than the flux of the weakest source,
$S_{weak}$, $S_{add}^{i}<S_{weak}$ for all non-resolved sources $i$. The total flux not considered in the
calculation is thus 
\begin{equation}
S_{tot}^{add}=\int \frac{dN}{dS}\,S\,dS <S_{weak}\cdot N_{add}\,.
\end{equation}
Here, $N_{add}=N_{tot}-N_{cat}$ is the number of additional sources which is
calculated from the total number of sources expected in the sky, $N_{tot}$,
subtracting the number of sources in the catalog, $N_{cat}$. Then, $\xi$ is calculated to
\begin{eqnarray}
\xi&=&\frac{S_{tot}^{cat}+S_{tot}^{add}}{S_{tot}^{cat}}\\
\xi_{\max}&=&\frac{S_{tot}^{cat}+S_{weak}\cdot
N_{add}}{S_{tot}^{cat}}\,,
\label{ximax}
\end{eqnarray}
where $S_{tot}^{cat}$ is the total flux in the catalog. 

\begin{figure}[h!]
\centering{
\includegraphics[width=10cm]{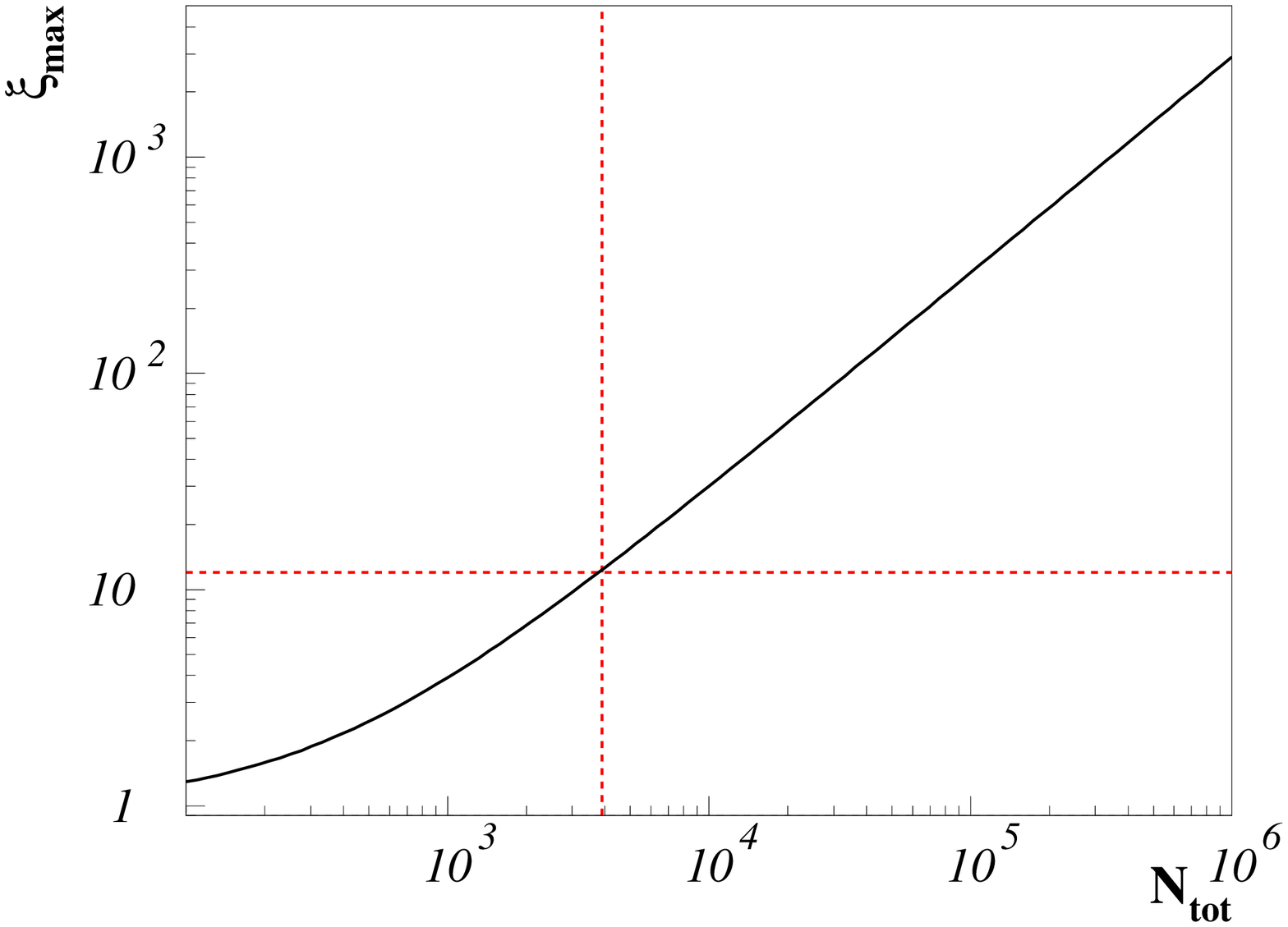}
\hangcaption[Maximum diffusive factor $\xi_{\max}$]{Maximum diffusive factor $\xi_{\max}$ versus total number of
  $>100$~MeV-emitting sources for {\sc EGRET} sources. With $\xi_{\max}=12$, the lower
  limit of the total number of sources contributing to the diffuse {\sc EGRET}
  background is $N_{tot}>3.9\cdot 10^3$.}
\label{ximax_ntot}
}
\end{figure}

Figure~\ref{ximax_ntot} shows the behavior of the maximum value of the
diffusive factor $\xi_{\max}$ according to
Equ.~(\ref{ximax}) at the example of {\sc EGRET} sources. The figure emphasizes the challenge in such a
representation. 
It is essential to have an estimate of the total number of
sources in the class to get a good estimate for $\xi_{\max}$. This implies the
necessity of the knowledge of the absolute lower luminosity limit. For FR-II
galaxies, this is relatively simple, since the number of sources decreases for
low luminosities~\citep{willott}. FR-I galaxies on the other hand have a high number of low
luminosity sources and it is not known at which luminosity the function
turns. 

In such a case, the diffusive factor can be quite large, since a high
contribution from low luminosity sources is expected. Thus, $\xi$ has 
explicitly been determined according to the neutrino flux model which is tested
by the limits. For FR-II galaxies, the investigated neutrino flux model
presented by~\cite{bbr_05} uses the luminosity
function by~\cite{willott}, labeled ''FR-II'' in Fig.~\ref{lumi_fcts} . In this case, a total number of $\sim 10^{5}$ sources
is expected and the maximum diffusive factor is determined to
$\xi_{\max}=160$. For the remaining source classes, FR-I galaxies,
GPS/CSS and QSOs, the value of $\xi_{\max}$ will not be given, since no
explicit neutrino model for these source classes is examined here. Note that
$\xi_{\max}$ gives an absolute upper limit: every source not included in the
stacking analysis is assumed to have the flux of the weakest included source,
$S_{weak}$. This results in an overestimate of $\xi$, since it is likely that
only a small fraction of the sources has such a high flux.

For {\sc EGRET} sources, Equ.~(\ref{ximax}) can be used to determine a lower
limit of sources contributing to the diffuse {\sc EGRET}
background. Figure~\ref{ximax_ntot} shows the dependence of the maximum value of
$\xi$ on the total number of sources in the source class of blazars emitting
at $>100$~MeV. Using a diffusive factor of $\xi=12$, the total number of
sources contributing to the total diffuse {\sc EGRET} background is
$N_{tot}^{EGRET}>3.9\cdot 10^{3}$.

\subsection{Comparing stacking diffuse limits with overall diffuse results}
\begin{figure}[h!]
\centering{
\includegraphics[width=10cm]{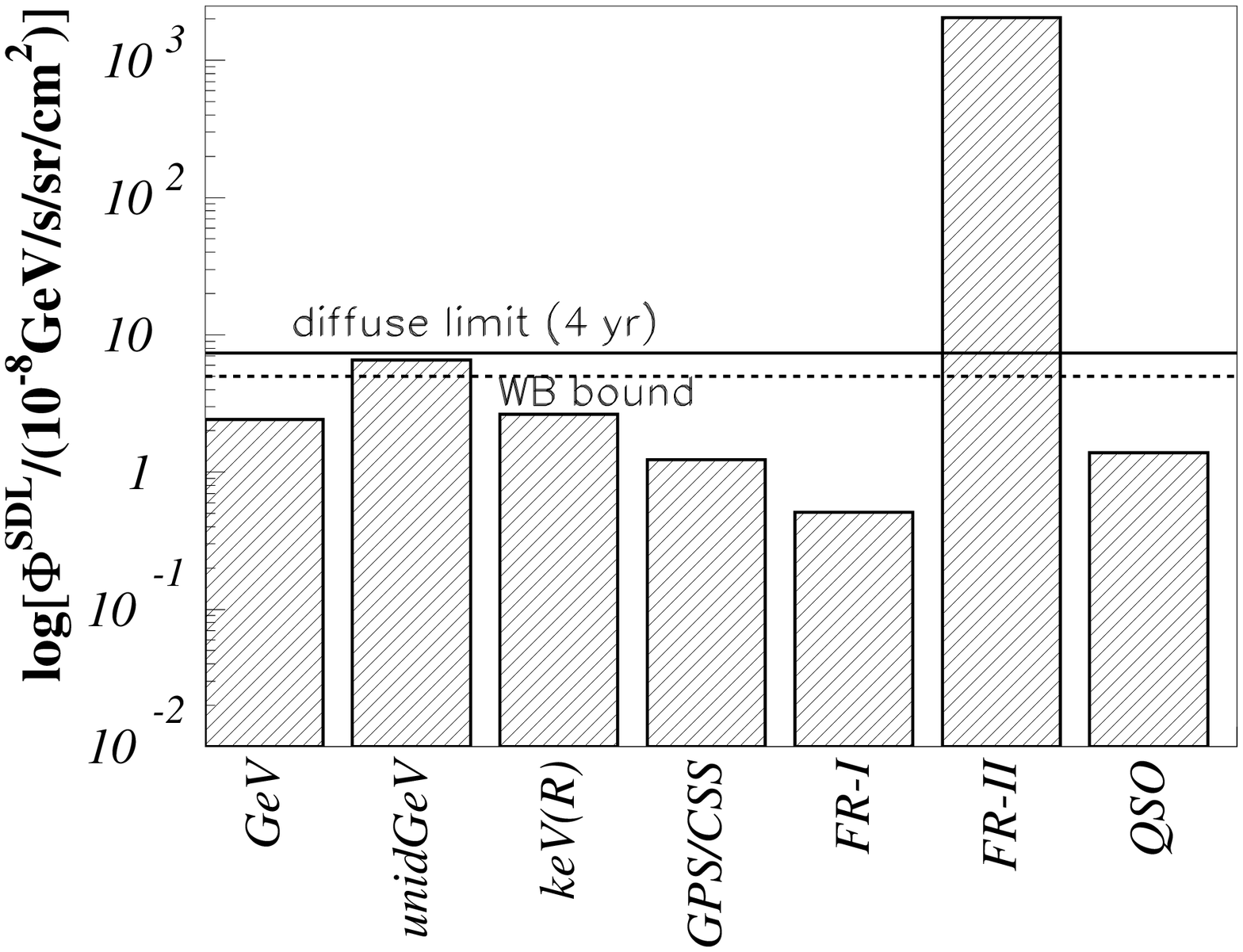}
\hangcaption{Diffuse limits of seven different AGN classes. The solid
  line represents the total diffuse limit of {\sc AMANDA}~\citep{diffuse_4yrs}. Several
  of the stacking diffuse limits are more restrictive than the general
  diffuse limit. The dashed line indicates the Waxman/Bahcall (WB) bound which
  gives a theoretical estimation of the maximum contribution of AGN to the
  diffuse neutrino flux.}
\label{stack_diff_5yr}
}
\end{figure}
Figure~\ref{stack_diff_5yr} shows the diffuse limits that could be derived
from the stacking approach for
each source class. It is compared to the diffuse limit of {\sc AMANDA} (solid
line) for the operation time considered (2000-2003),\\ $\dl=7.4\cdot
10^{-8}\,\diffunits$, given in~\citep{diffuse_4yrs}.
Several stacking diffuse limits are more restrictive than the general diffuse
limits. 
The
exact numbers are given in Table~\ref{5yr_diff_tab}, where the stacking limit,
the diffuse bound derived from the stacking method and the ratio of the latter
and the general diffuse limit are compared. Note that in the case of CSS/GPS,
QSOs and FR-I galaxies, the diffusive factor still has to be applied.

Figure~\ref{stack_diff_5yr} also indicates the Waxman/Bahcall bound
(WB bound, dashed
line). This bound gives a theoretical estimate of the maximum contribution to
be expected in neutrinos from Active Galactic Nuclei~\citep{waxbah}. This
bound is valid for optically thin sources at an energy of $\en\sim 10^{17}$~eV, where it
is most restrictive. The
description for all energies is given by~\cite{mpr}. Five of the seven source class limits lie below
the indicated WB bound. This shows that the sensitivity reached with this
method in {\sc AMANDA} is already extremely high. Next generation's neutrino telescopes such as
{\sc IceCube} and {\sc KM3NeT} which will have a cubic kilometer of ice resp.~water instrumented will therefore
be able to provide more information either by giving very strong restrictions
on prevailing models or by confirming a positive neutrino signal.   

Note that the class of TeV blazars is on the other hand an optimal candidate
for a general diffuse search. The contribution of photon-observable TeV blazars
is very small, since many sources are hidden in photons due to the strong
absorption, while they would still be visible in neutrinos. This will be
discussed in Section~\ref{compare}.

\begin{table}
\centering{
\begin{tabular}{l|lll|l}
Source class&$\stl$&$\sdl$&$\dl/\sdl$&models\\\hline\hline
{\sc EGRET} GeV&2.71&7.25&1.02&\citep{mpr},\\
           &   &&&  \citep{mannheimjet},\\
         &     &   &    & \citep{muecke}\\
unid. {\sc EGRET}&31.7&6.56&1.23&-\\
infrared &10.6&EXCL.:&too few&-\\
keV ({\sc HEAO-A})&3.55&EXCL.:&too few&-\\
keV ({\sc ROSAT})&9.71&2.65&2.79&\citep{stecker96},\\
                 &    &    &    &\citep{nellen},\\
                 &    &    &    &\citep{alvarez_xrays}\\
TeV blazars&5.53&EXCL.:&too few$^{*}$&\citep{mpr},\\
           &    &      &             &\citep{muecke},\\
           &   &         &                    &\citep{mannheimjet}\\
CSS/GPS&5.94&$1.23\cdot\xi_{model}$&$6.02/\xi_{model}$&-\\
FR-I incl.~M-87&4.11&EXCL.: M87&dominant&-\\
FR-I excl.~M-87&2.91&$0.51\cdot \xi_{model}$&$14.5/\xi_{model}$&-\\
FR-II&30.4&$2.05\cdot 10^{3}$&$3.61\cdot 10^{-3}$&\citep{bbr_05}\\
QSOs&6.70&$1.39\cdot\xi_{model}$&$5.34/\xi_{model}$&-\\
\end{tabular}
\caption{Table of the source class limits obtained with the stacking
  method. Five years of data, 2000-2004, have been used for the analysis with
  {\sc AMANDA} in \citep{5yrs}. The stacking limit is given in units of $10^{-8}\,\pointunits$
  while the stacking diffuse limits is given in units of
  $10^{-8}\,\diffunits$. 
\hspace{7cm}
$^*$While the class of TeV blazars cannot be used
  to determine a stacking diffuse limit, the general diffuse limit gives an
  upper limit to the contribution of TeV-observable blazars to the total diffuse
  flux as it is shown in Section~\ref{compare}.
\label{5yr_diff_tab}
}}
\end{table}
\clearpage
\section{Direct implications for AGN neutrino flux models \label{compare}}
The diffuse limits discussed in Sections~\ref{introduction} and \ref{stack_results}
constrain some of the currently discussed neutrino flux models. However,
it is known that these models bear different uncertainties in both
normalization and spectral shape due to a lack of knowledge of the conditions
at the source. 

In this section, the neutrino flux models having been discussed in Section~\ref{introduction} will be examined
with respect to the limits obtained.
\subsection{TeV blazars \label{tev}}
The detection of TeV photon emission from distant sources is limited due to
the absorption of high energy photons by the extragalactic background
light~\citep[e.g.][]{stecker92,tanja}.
The absorption factor
$\eta$ describes the ratio of the total emitted TeV photon flux from
HBLs and the TeV flux from HBLs up to a redshift $z_{\max}$. It is a
measure of the absorption of TeV photons. 
The ratio of a diffuse neutrino signal from photon-observable TeV blazars will be calculated
using the general diffuse limit and taking into account the absorption factor
for TeV photons: Only a relatively small fraction of all TeV
blazars can be identified in TeV photons, since sources at high redshifts are
hidden due to the strong absorption. 

In this paragraph, the absorption factor
and the general diffuse neutrino limit are used to derive the maximum
contribution of TeV photon-observable sources to the total diffuse neutrino flux.
\subsubsection{The absorption factor $\eta$}
In the case of TeV blazars, the absorption factor $\eta$ is much greater than unity due to the strong
absorption of TeV photons by the extragalactic background light. HBLs seem
to have no or even a slightly negative evolution~
\citep[e.g.][]{beckmann,blazar_ev,laurent}. 

For a no-evolution scenario as it is discussed for BL Lacs, the co-moving
density $\rho(z)$ is considered to be constant with redshift, $\rho(z)=\mbox{constant}$.
Using a negative evolution with less than $(1+z)^{-0.2}$ does not
change the results significantly. We also neglect the positive source evolution which
has to be present up to a certain redshift $z^*$: both effects positive
evolution up to $z^*$ and negative evolution at higher redshifts are believed
to cancel. Each effect for itself alters the result less than $10\%$ in
opposite directions. 

The normalization of the co-moving density is not important in this calculation,
since only ratios of concrete values of $\rho$ are considered and the constant
of proportionality cancels.

The neutrino flux from sources up to a certain redshift $z_{\max}$ is given by
\begin{equation}
\frac{dN}{dE}=\phi_0\cdot\int_{z=0}^{z_{\max}} E'(z)^{-2}\cdot \rho(z)\frac{dV_{c}}{dz}\,dz\,,
\end{equation}
where $E'(z)=(1+z)\cdot E$ is the energy of the neutrino at the source. The
$E^2$-weighted flux is thus given as
\begin{equation}
E^2\frac{dN}{dE}=\phi_0\cdot\zeta(z_{\max})\,.
\label{all}
\end{equation}
Here, $\zeta(z_{\max})=\int_{z=0}^{z_{\max}} (z+1)^{-2}\cdot
\rho(z)\,dV_{c}/dz\,dz$ is the evolution factor depending on the upper redshift
integration limit $z_{\max}$. If all present sources are considered,
$z_{\max}\approx 7$ is a reasonable value. Recent searches for luminous
galaxies at redshifts between $z=6-8$ imply that only few ultra-luminous
objects only exist beyond $z=7$~\citep{iye06,bouwens06}. At higher redshifts,
no large galaxies are observed, leading to the conclusion that only smaller galaxies
can be present. Also, the
results to not change significantly when going up to higher redshifts: the
major contribution comes from redshifts of $z<3$.

\begin{figure}
\centering{
\includegraphics[width=10cm]{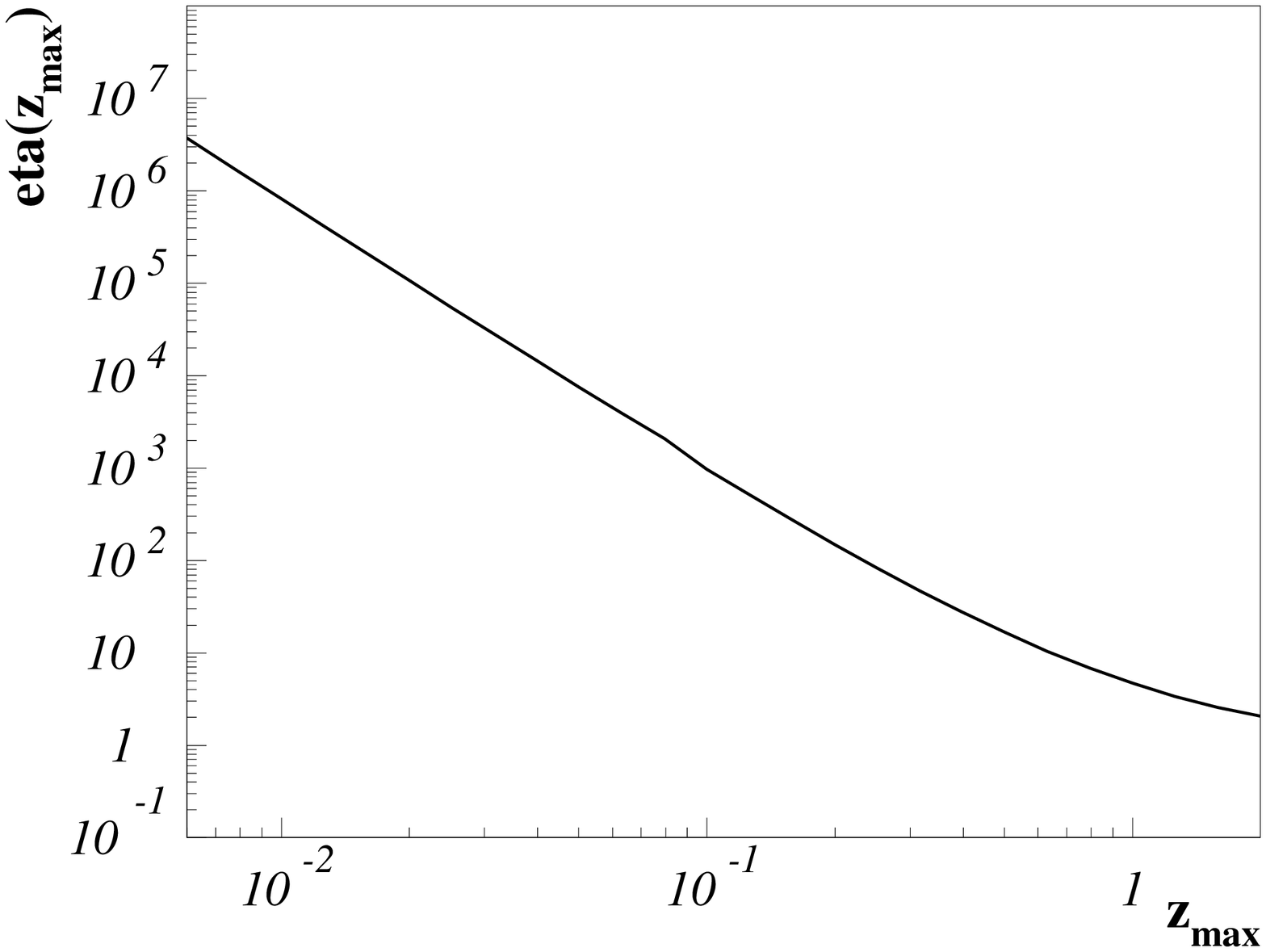}
\hangcaption{Absorption factor $\eta$ versus maximum redshift $z_{\max}$. For TeV
  sources, the most distant
source observed is 1ES1101-232 at z=0.186~\citep{hess_blazar}. 
$\eta(z_{\max}=0.3)=54$ is used for TeV
sources.}
\label{absorption}
}
\end{figure}

The absorption factor $\eta$ is given by the ratio of the total flux and the
contribution that is
observed by TeV-photon experiments:
\begin{equation}
\eta(z_{\max})=\frac{\zeta(7)}{\zeta(z_{\max})}\,.
\end{equation}
Here, $z_{\max}$ is the maximum redshift at which TeV photon sources can be
identified by present Imaging Air Cherenkov Telescopes (IACTs).
TeV sources have up to today been observed up to a redshift of z=0.186 - 1ES1101-232 \citep{hess_blazar}. As a conservative upper limit, the maximum redshift of
observation with IACTs is taken to be $z_{\max}=0.3$, since an upper
bound is derived from these values. The absorption factor depending on
$z_{\max}$ is shown in Fig.~\ref{absorption}. At $z_{\max}=0.3$, the
numerical value is
\begin{equation}
\eta(z_{\max}=0.3)=54\,.
\end{equation}

It is important to note that for a lower energy threshold (e.g.~30~GeV), IACTs could be able to detect sources up to $z_{\max}\sim
1$. This would lead to a much lower absorption factor of
\begin{equation}
\eta(z_{\max}=1)=4.6\,.
\end{equation}

\subsubsection{An upper limit to the contribution from photon-observable TeV blazars}
Using the above reflections, one can argue that a diffuse limit can be
used to estimate the maximum diffuse signal from HBLs resolved
in TeV photons: 
From the existing general diffuse limit, it is known that no source
type and, thus particularly also not the class of TeV blazars, can contribute more than $\dl=7.4\cdot 10^{-8}\,\diffunits$:
\begin{equation}
E^2\left.{\frac{dN}{dE}}\right|_{TeV}<\dl\,.
\label{all_dl}
\end{equation}
An upper limit to the normalization constant $\phi_0$ of the neutrino flux can
be derived by inserting Equ.~(\ref{all}) with $z_{\max}=10$ in Equ.~(\ref{all_dl}): 
\begin{equation}
\phi_{0}<\frac{\dl}{\zeta(10)}\,.
\end{equation}
The contribution of sources at $z<z_{\max}$ is
given as
\begin{equation}
E^2\left.{\frac{dN}{dE}}\right|_{TeV}(z_{\max})=\phi_0\cdot\zeta(z_{\max})\nonumber<\frac{\zeta(z_{\max})}{\zeta(10)}\dl
\end{equation}
with the final result of
\begin{equation}
E^2\left.{\frac{dN}{dE}}\right|_{TeV}(z_{\max})<\eta(z_{\max})^{-1}\cdot \dl\,.
\end{equation}

\begin{figure}[h!]
\centering{
\includegraphics[width=10cm]{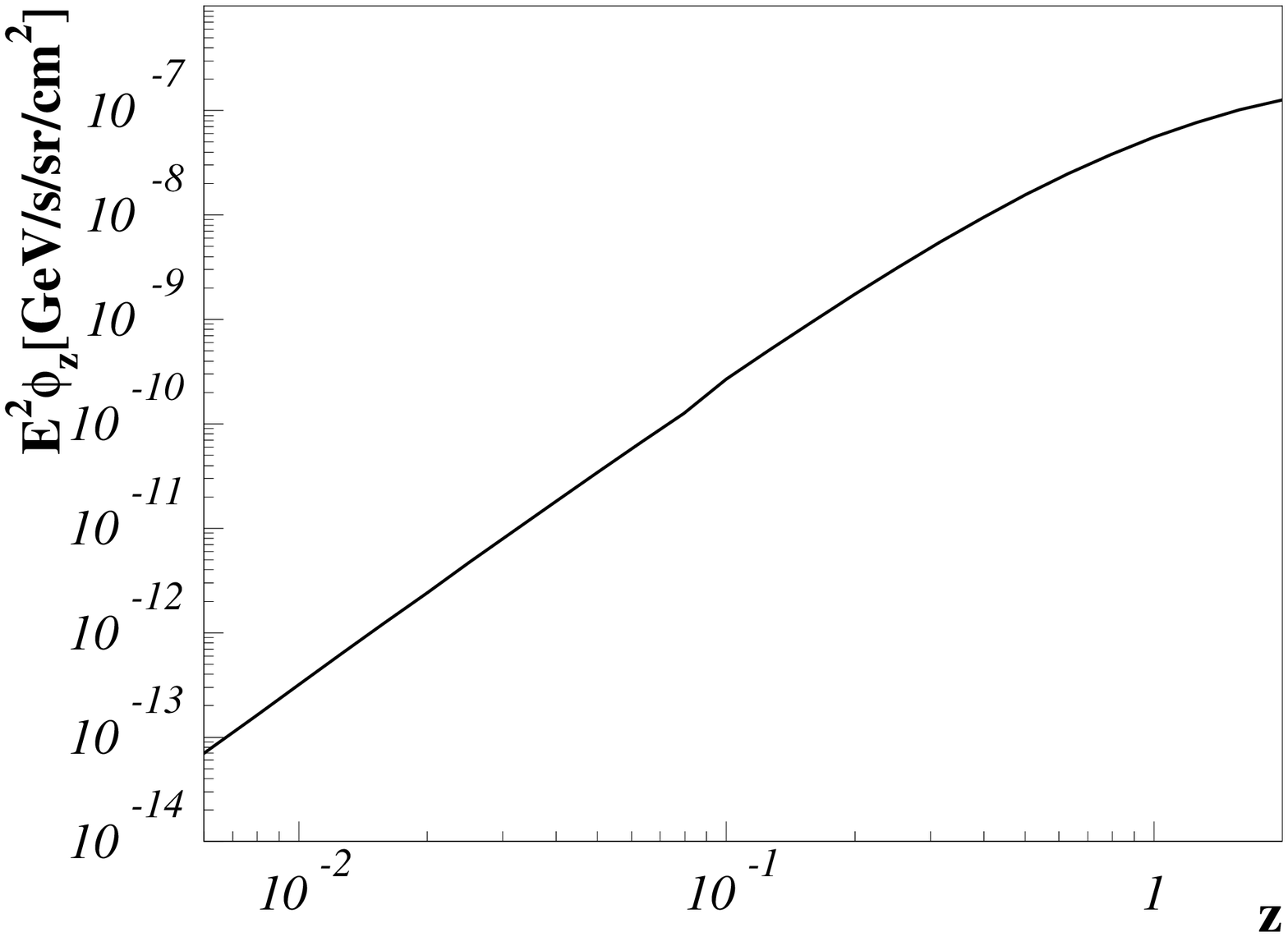}
\hangcaption{Upper limit on the contribution of photon-observable TeV blazars for
  the no-evolution scenario for HBLs.}
\label{flux_fraction}
}
\end{figure}

The upper limit on the neutrino flux up to $z_{\max}$  is shown in Fig.~\ref{flux_fraction}.
The curve shows the no-evolution scenario. 

\begin{figure}[h!]
\centering{
\includegraphics[width=10cm]{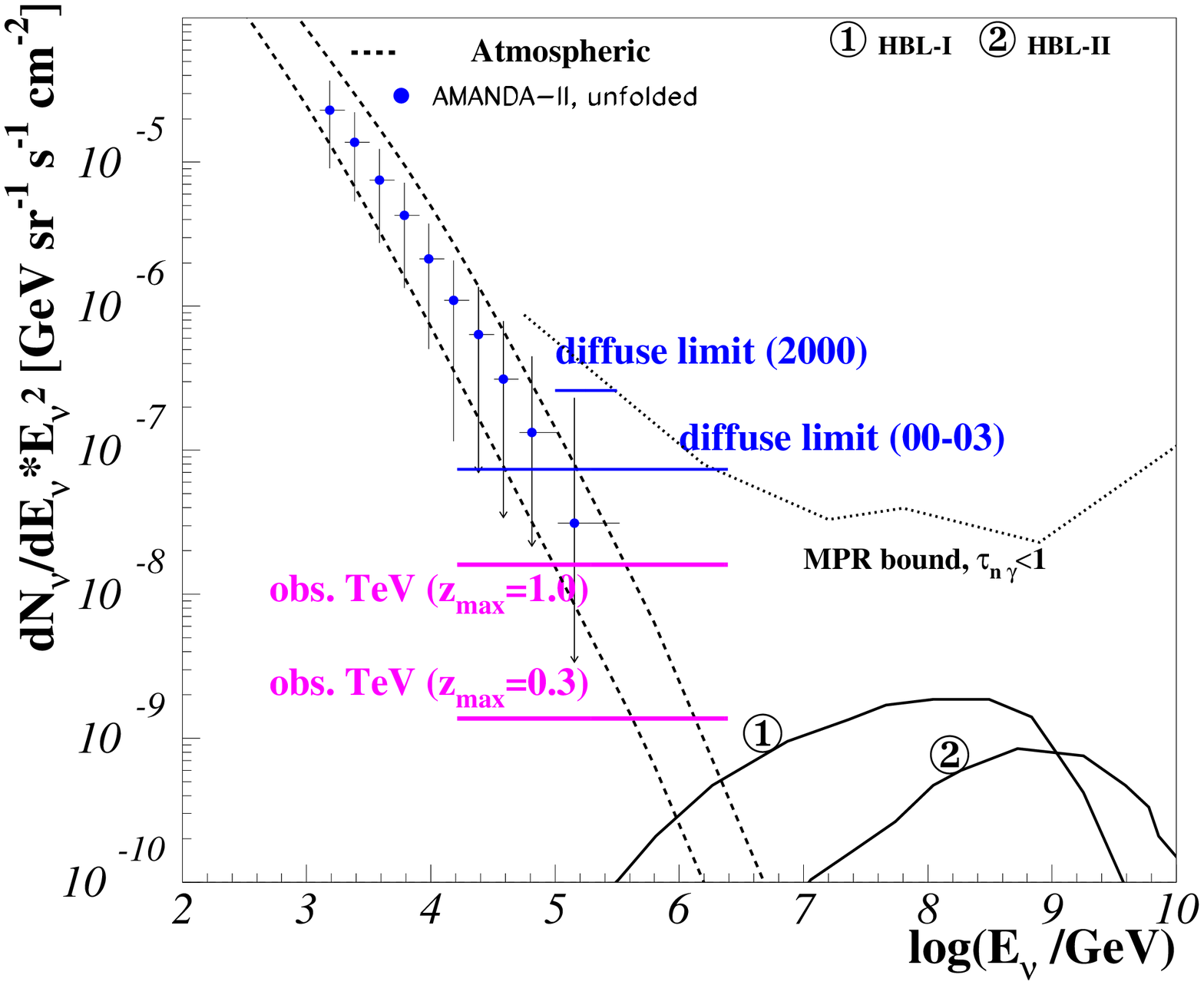}
\hangcaption{Neutrino predictions for optically thin sources. TeV photons from
  decaying pions escape together with neutrons which make up charged CRs by decaying in the vicinity
of the source in protons. All models have been corrected for oscillations. The
  {\em MPR-bound} derivation is described
in~\citep{mpr}. Predictions for the contribution from HBL sources ({\em HBL-I}
  and {\em HBL-II}) depend
on the extension of the acceleration region and the jet frame target photon
density. Model 1 and 2 represent two different parameter settings,
see~\citep{muecke}. The general diffuse limit is shown as well as the limit to
  photon-observable TeV blazars ({\em obs.TeV}), indicating that the major contribution to the
  diffuse neutrino signal from TeV blazars is expected to come from hidden TeV
  photon sources for two different maximum redshifts, $z=0.3$ and $z=1$.}
\label{blazars_thin}
}
\end{figure}

The contribution of photon-observable TeV blazars to the diffuse
neutrino flux is therefore limited to 
\begin{equation}
E^2\left.{\frac{dN}{dE}}\right|_{TeV}(z_{\max}=0.3)<1.37\cdot10^{-9}\,\diffunits\,, 
\end{equation}
where $\eta(z_{\max}=0.3)=54$ is used as a lower limit.
The contribution of sources observed by IACTs
is thus about three orders of magnitude lower than the possible total
contribution. This is displayed in Fig.~\ref{blazars_thin} together with the general
diffuse limit. The limit to the maximum contribution from
photon-observable TeV blazars is shown ({\em obs.~TeV}) as well. The contribution is a
factor of $\eta^{-1}\approx0.019$ lower than the diffuse
limit. This indicates that a diffuse analysis of TeV sources is most effective
with an overall diffuse approach, since most of the TeV sources are hidden due to the
strong absorption at such high photon energies. The stacking of the strongest
sources in the sky in comparison only comprise a small fraction of the total
diffuse flux as these calculations show. This result stands in contrast to the other sources
samples where the selection of the strongest sources yields stronger
restrictions than a general search for a diffuse signal, see following paragraphs.

The importance of reaching down to low energies with IACTs
is demonstrated in this calculation, since lowering the energy threshold to
observe sources up to $z=1$ would already include sources with a
one order of magnitude higher maximum neutrino flux of
\begin{equation}
E^2\left.{\frac{dN}{dE}}\right|_{TeV}(z_{\max})<1.60\cdot10^{-8}\,\diffunits\,. 
\end{equation}
\subsection{Optically thick cases: sources of MeV and GeV $\gamma$ emission\label{thick}}
If the sources are optically thick to photon-neutron interactions,
$\tau_{\gamma\,n}>>1$, the photon signal which is emitted from the sources
lies in the MeV to GeV range. Thus, {\sc EGRET} and {\sc COMPTEL} diffuse cosmic photon fluxes
are used to determine the expected neutrino contribution. 
\subsubsection{EGRET blazars}
\begin{figure}[h!]
\centering{
\includegraphics[width=10cm]{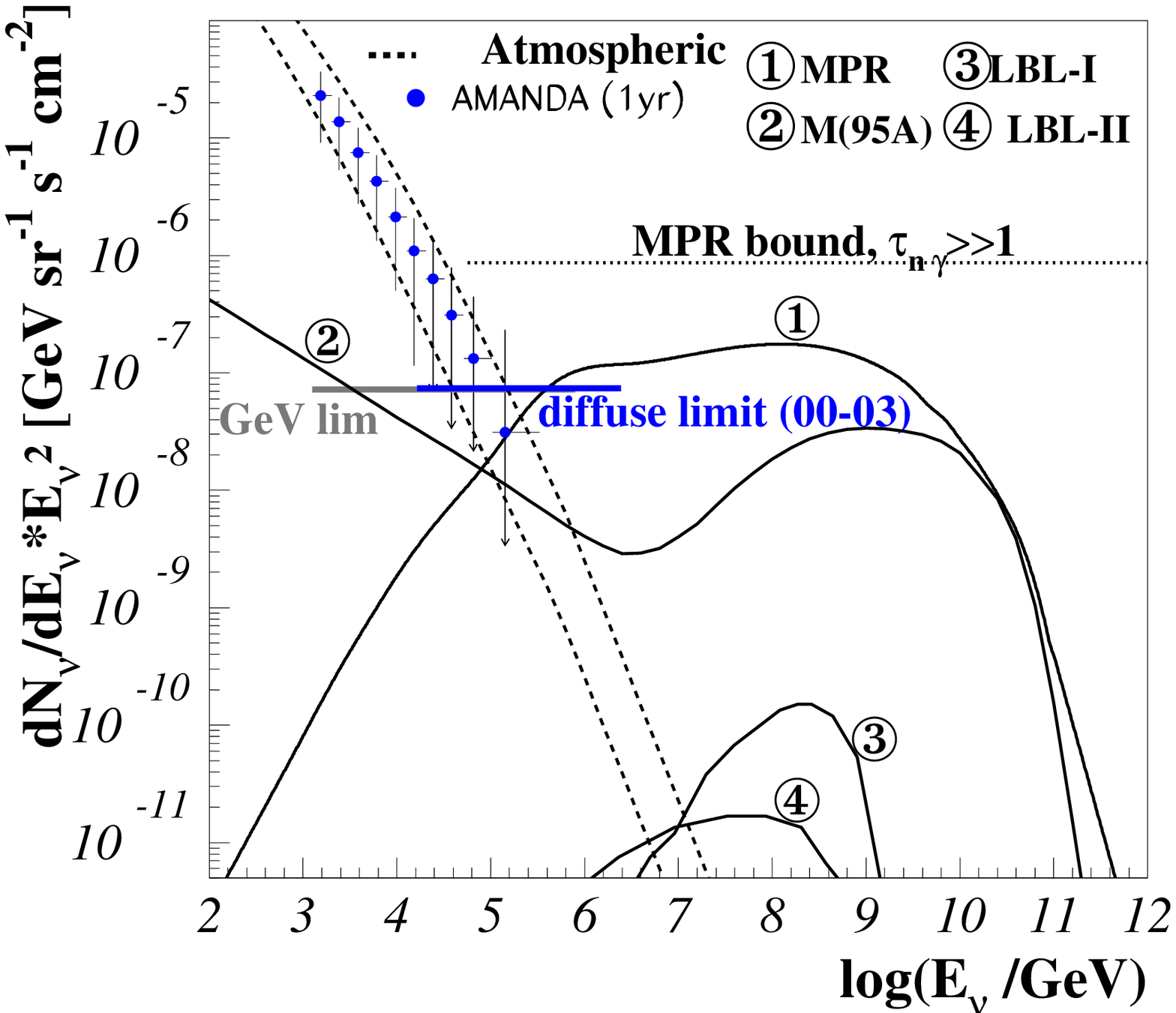}
\hangcaption{{Neutrino flux predictions for optically thick sources, normalized to
  the diffuse {\sc EGRET} flux above $100$~MeV. Model 1 ({\em MPR}) represents the maximum
  contribution from blazars within the framework of~\cite{mpr} in which the
  {\em MPR bound} for optically thick sources (dashed line) is derived. Model
  2 ({\em M95,A}) - gives the neutrino spectrum from $p\,p$ and $p\,\gamma$ interactions in
  blazars determined in~\cite{mannheimjet}. Models 3 and 4  ({\em LBL-I}
  and {\em LBL-II}) represent the
  prediction of neutrinos from LBLs given two different parameter settings,
  see~\cite{muecke}. The stacking diffuse limit for the class of identified
  {\sc EGRET} sources is indicated with
  $\xi=\xi_{\max}=12$ where it is assumed that AGN produce the complete diffuse
  {\sc EGRET} signal. 
\label{blazars_egret}
}}
}
\end{figure}

Figure~\ref{blazars_egret} shows neutrino flux predictions in connection with
{\sc EGRET} data and the derived stacking diffuse limits for a neutrino signal from
{\sc EGRET} sources. Since the indicated models are all normalized to the {\sc EGRET} diffuse spectrum,
the limit of identified {\sc EGRET} blazars applies. 
The gray line between $1$~TeV and $10^{2.8}$~TeV indicates the most conservative
calculation where it is assumed that the total diffuse background as measured
by {\sc EGRET} it produced by AGN $(\xi=12)$. 
The limit falls short of the MPR bound and also touches
the maximum prediction for the source class predicted by MPR and
it is possible to reach sensitivities at the level
of predictions concerning a neutrino signal from blazars. The same is valid
for the prediction by Mannheim (M(95), A): The proton-proton contribution from
optically thick blazars can be constrained even
though the atmospheric flux exceeds the prediction at these energies. With an
overall diffuse analysis, it would not be possible to extract any information
on this low-energy part of the spectrum from AGN. 

The contribution from LBL within the
proton-blazar model dominates the total observable flux at ultra high
energies. With the current data, it is not possible to get to any reliable
conclusion from the stacking method, since it is valid at energies $E<$~PeV
while the flux is present at much higher energies. A stacking analysis of cascade events with
second generation telescopes such as {\sc IceCube} would be an option to explore
regions of energies at $>$~PeV. A stacking approach for cascades is however accompanied
by the challenge of getting a reasonable directional reconstruction of the neutrino events,
since the cascade signal is generally not as boosted as a signal from
neutrino-induced muons.

Even in the most conservative case of a diffusive factor of $\xi_{\max}=12$, the
model of proton-proton energies can be constrained at low energies. With the
stacking method, the diffuse limit can be extended to sensitivities far below the
flux of atmospheric neutrinos.
\subsubsection{{\sc COMPTEL} blazars}
A stacking analysis using {\sc COMPTEL} blazars has not been done yet. The option is
discussed in Section~\ref{class_analysis}. Current diffuse neutrino flux
limits are still about a factor of 2 or more above the predictions. The
hypothesis of an avalanched TeV signal down to MeV energies will be tested by
{\sc IceCube}. 
\subsection{The diffuse X-ray background and Radio Weak AGN}
The X-ray component of radio weak AGN can be correlated to the emission of
neutrinos at the foot of the jet. This has been investigated in calculations
of~\cite{nellen}, \cite{stecker96} as well as~\cite{alvarez_xrays}. All models are constrained by the
current general diffuse {\sc AMANDA}
limit. The resolved {\sc ROSAT} sources which have been used to determine the
stacking limit are radio loud and thus, the limit is not as easily applicable to these
predictions. If the same production mechanism is assumed in radio loud as in
radio weak sources, the stacking diffuse limit of {\sc ROSAT} sources would apply in
this case. 

However, the limit to radio weak objects ${\sdl}_{rq}$ is about a factor of 10
higher than the calculated limit for radio loud sources ${\sdl}_{rl}$, since
radio loud objects are about a factor of 10 less frequent than radio weak sources,
\[
{\sdl}_{rq}\lesssim 10\cdot {\sdl}_{rl}=1.72\cdot10^{-7}\diffunits \,.
\]
The more restrictive limit in this case is therefore the general diffuse limit.

\begin{figure}[h!]
\centering{
\includegraphics[width=10cm]{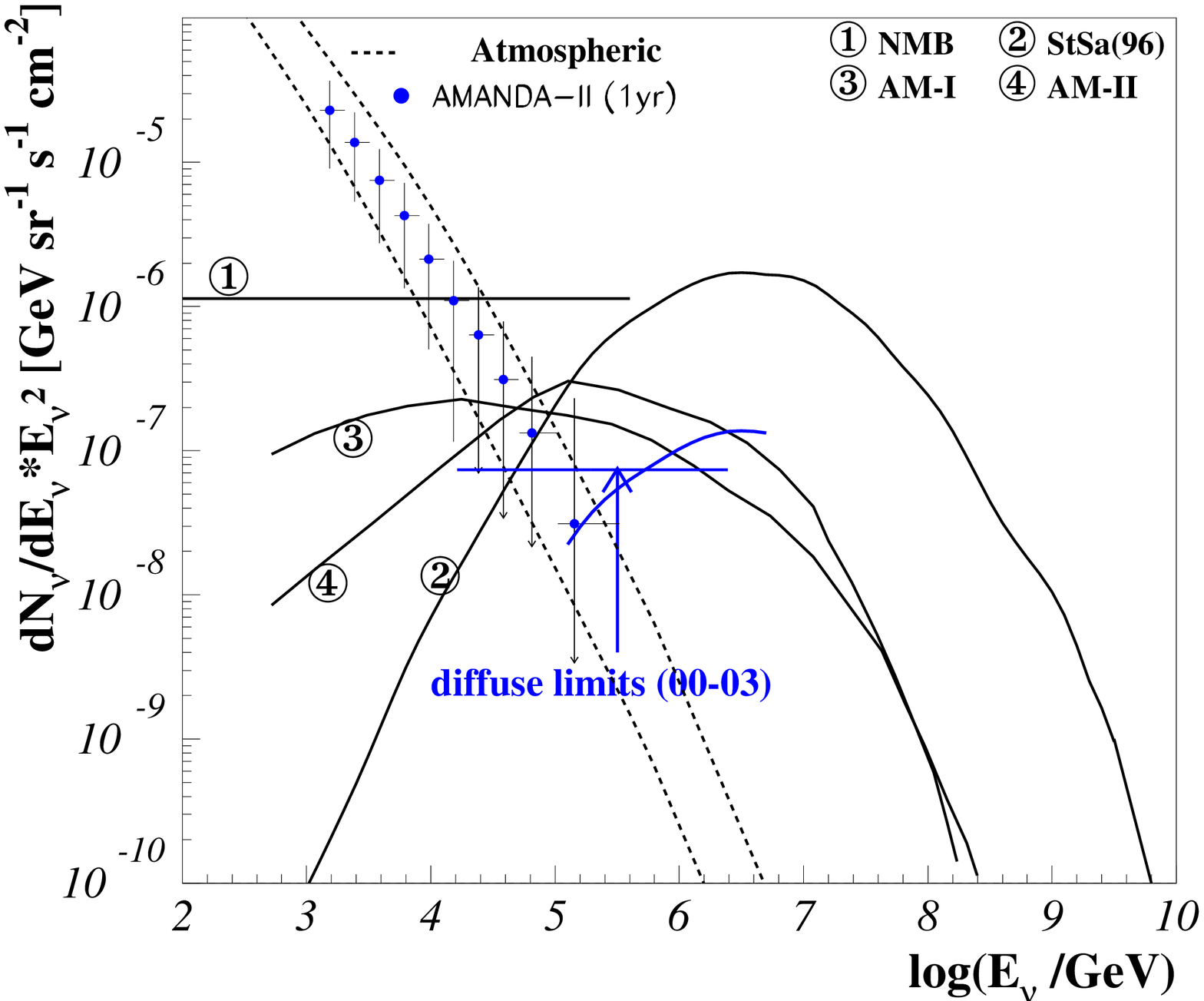}
\hangcaption{\cite{stecker96} ({\em StSa96}, model 2) predict a large contribution of neutrinos
  from radio weak sources using {\sc ROSAT} data to normalize the
  spectrum. \cite{nellen} ({\em NMB}, model 1) use the same normalization
  method. Model 3 and 4 ({\em AM-I} and {\em AM-II}, \citep{alvarez_xrays}) use the
  luminosity function of radio quiet AGN and normalize the single source
  spectrum. While the black hole mass is varied in model 3, model 4 varies the
  accretion rate. All these models have been corrected for neutrino oscillations. All models are
  excluded in their published form by {\sc AMANDA}'s diffuse limit. The
  typical $E^{-2}$ limit is represented by the blue horizontal line. The
  curved blue line, crossing the $E^{-2}$-limit represents the {\sc AMANDA}
  limit for the years 2000-2003 calculated for a flux shaped as the model by
  \cite{stecker96} and \cite{stecker_mod}.}
\label{rosat}
}
\end{figure}

Figure~\ref{rosat} shows the prediction of neutrinos being produced
coincidentally with X-rays at the foot of AGN jets. The diffuse {\sc ROSAT}-measured
background has been used in two models \citep{nellen,stecker96} to normalize
the neutrino spectrum. In the third case presented by \cite{alvarez_xrays}, the luminosity evolution function of
radio quiet AGN has been used so that the result also applies to the
correlation with the X-ray diffuse background. 

In the case of the prediction by \cite{nellen}, {\sc AMANDA}'s diffuse neutrino flux limit lies an order of
magnitude below the flux. Stecker et al.~updated their calculation by
normalizing to the diffuse flux as measured by
{\sc COMPTEL}~\citep{stecker_mod}. This is discussed
in the context of the MeV photon emission in correlation with neutrinos in Section~\ref{thick}. The
absence of neutrinos from {\sc ROSAT}-detected sources implies strongly that the X-ray emission from AGN cannot be directly
correlated to neutrino emission. While the missing emission of neutrinos
 disfavors a hadronic scenario for the X-ray emission from AGN, it is consistent with Inverse Compton (IC)
models for the X-ray emission and underlines leptonic models for the
production of X-rays in AGN. 

Calculations by \cite{stecker96} show a flux
peaking at higher energies compared to the prediction from~\cite{nellen}. The main reason is that
\cite{nellen} use a more conservative estimate of the maximum energy and a
simpler approach for the spectral behavior. The spectrum of the model
by~\cite{stecker96} is not exactly $E^{-2}$-shaped, while the
{\sc AMANDA} limits are typically derived assuming an $E^{-2}$ spectrum. The modeling
of different spectral shapes, including~\cite{stecker96}, has been done
recently, see~\citep{jess_madison06}. The curved blue line indicates the limit
modeled according to the model of~\cite{stecker96}. The limit lies half an
order of magnitude below the prediction.

While the previously described models normalize the diffuse spectrum by using
{\sc ROSAT} data, calculations by~\cite{alvarez_xrays} use the radio luminosity
function for radio quiet AGN and a single source normalization. Further
differences with respect to model 1 and 2 are the maximum energy and the break
in the neutrino spectral shape. The underlying idea is the same by assuming
proton-proton and proton-photon interactions at the foot of AGN jets, leading
to X-rays in coincidence with neutrinos. This model overproduces neutrinos as
well as the previous ones. 

Three independent calculations on the correlation of X-ray emission from radio
quiet AGN and neutrino emission have been examined. The absence of neutrino emission from the foot of relativistic jets 
implies that the particles there are not accelerated to high energy, and 
then interact.  However, the concept that the innermost ring of the 
accretion disk, just underneath the jet, turns into an Advection 
Dominated Accretion Flow (ADAF) with a very high temperature
\citep[e.g.][]{falcke,cr5,rmb,donea,mahadevan}, is consistent with the results. 
This implies that the spin 
parameter of the black hole is high, above 0.95, and that hadronic 
interaction at weakly relativistic temperatures produce charged pions, 
which decay and provide an energetic particle seed population for further
acceleration downstream for the radio 
emission of the jet~\citep{krishna}.  This is then consistent with 
an Inverse Compton explanation of the X-ray emission~\citep{msr}, and predicts that there should be a large production of energetic 
neutrinos at an energy commensurate with the pion mass.
\subsection{Radio galaxies}
The radio emission of Active Galactic Nuclei is likely to be directly
correlated to neutrino emission in the jet. This has been discussed
by~\cite{bbr_05} at the example of FR-II radio galaxies and flat spectrum
radio quasars (FSRQs). 
In both cases, the normalization of the neutrino spectrum depends on many different intrinsic factors, i.e.~the
correlation between radio and disk luminosity and the parameterization of the
maximum proton acceleration energy. The jet-disk correlation of AGN has been worked out
by~\cite{falcke,falcke1,falcke3}. Although the model includes several parameters,
all are fixed by the comparison with the data except the accretion rate. Within that model, it is
shown that the parameter reach extreme values and are not strongly
scattered. The jet-disk symbiosis model has been proven for different source types, reaching from
microquasars to quasars.

A parameter in the calculation of the neutrino flux is the optical
depth of the source which is unknown. In~\citep{bbr_05}, the effective optical depth of the source,
$\tau_{eff}$ is defined as product of the proton-photon optical depth
$\tau_{p\,\gamma}$
and a reduction factor due to contributions from Bethe-Heitler production $\eta'$,
\begin{equation}
\tau_{eff}=\tau_{p\,\gamma}\cdot \eta'\,.
\end{equation}
It can be derived from the solution of the transport equation, the neutrino flux is proportional to
$\tau_{eff}/(1-\exp(-\tau_{eff}))$. For large depths, the neutrino flux is
  therefor proportional to $\tau_{eff}$.
The effective optical depth has been chosen to be $\tau_{eff}=1$ in the
calculations described above. The optical depth of proton-photon interactions
is determined by the product of the photo-hadronic cross section
$\sigma_{p\,\gamma}$ and the photon density in the source, $n_{\gamma}$. These
in turn depend on the disk luminosity $L_{disk}$ and the extent $r$ of the source
such as
\begin{equation}
\tau_{p\gamma}=800\frac{L_{46}}{r_{17}}\,,
\end{equation}
with $L_{46}:=L_{disk}/10^{46}$~erg/s and $r_{17}:=r/10^{17}$~cm. This shows that
the photo-hadronic optical depth itself uncertain, since low values around
unity are possible as well as extremely high numbers. {\sc AMANDA} limits will be used to derive limits
on the optical depth of FR-II galaxies and FSRQs.

Note that the determination of $\tau_{eff}$ happens only within the specific
model of neutrino production as it is described above. The difficulty of
drawing more general conclusions lies in the uncertainty of the spectral
index. This highly depends on the original spectral behavior of the
protons. This can only be simulated and is not directly observable.

\subsubsection{FR-II radio galaxies}
Using a generic $E^{-2}$ spectrum for the neutrino flux from FR-II galaxies, the
flux normalization has been derived to be
\begin{equation}
\Phi=1.43\cdot 10^{-7}\,\diffunits\,.
\label{bbr_norm}
\end{equation} 

Comparing the normalization in Equ.~(\ref{bbr_norm}) with the diffuse
limit shows that a flux with the chosen parameter settings is not detected. An upper limit to the optical
depth can be derived from the limit yielding
\begin{equation}
\tau_{eff}<0.5\,.
\end{equation}
Within three years
of {\sc IceCube}, it will be possible to explore sources with $\tau_{eff}>0.029$.

\begin{figure}
\centering{
\includegraphics[width=10cm]{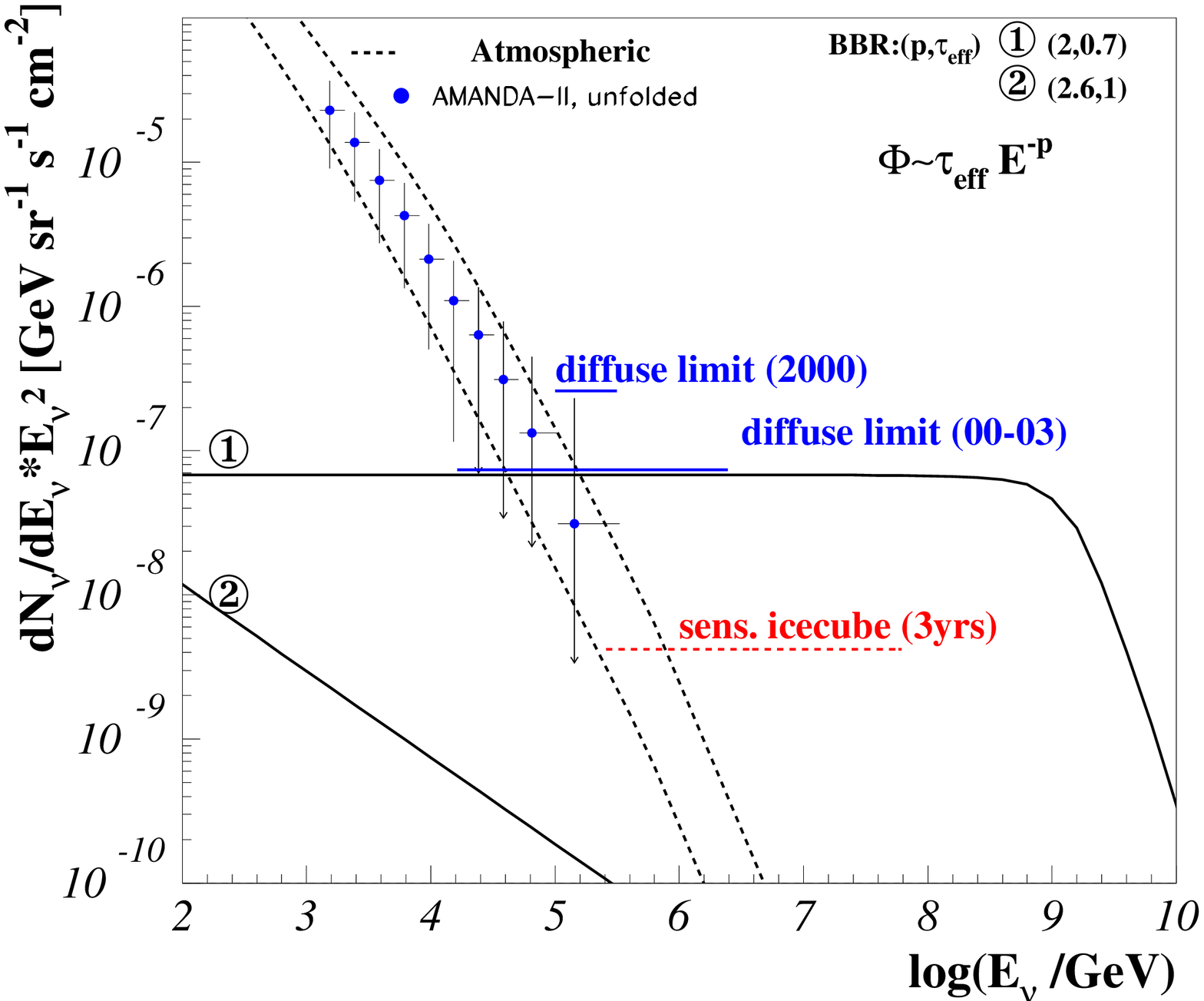}
\hangcaption{Contribution to the diffuse neutrino background from FR-II radio
  galaxies. For an $E^{-2}$ spectrum, it is constrained to an optical depth of $\tau_{eff}<0.5$.
A steeper
spectrum $E^{-2.6}$ (model 2) is consistent with the current limit.}
\label{fr2}
}
\end{figure}

Figure~\ref{fr2} shows the the prediction of
a neutrino flux from FR-II radio galaxies. Model (1) is the original
calculation of~\cite{bbr_05} modified by a factor $\tau_{eff}$=0.5. This is
the maximum contribution for an $E^{-2}$ signal from those sources not violating the diffuse limit of
{\sc AMANDA}. However, there is
another production scenario which leads to a significant reduction of a
contribution from FR-II galaxies. The synchrotron spectral index of FR-II galaxies $\alpha$ is correlated to the
neutrino spectral index $p$ by $p=2\cdot \alpha+1$~\citep[e.g.]{rybicki}. The mean synchrotron spectral
index is $\alpha\approx 0.8$ for large scale emission and therefore, the neutrino spectrum can be as
steep as $E^{-2.6}$. This prediction is shown in Fig.~\ref{fr2} as model
(2). In that case, the contribution is very low and does neither violate the
current diffuse limit, nor is it within reach of {\sc IceCube}'s sensitivity for
three (or even five) years of operation~\citep{diffuse_4yrs}.
In such a case, an observation of a neutrino signal
from FR-II galaxies would be very difficult. If the spectrum 
were this steep for real, then all Ultra High Energy Cosmic Ray (UHECR)
arguments would fail and so it may be more appropriate to assume a flatter
spectrum. The radio spectral index of the hot spots are usually about
$\alpha\sim0.6$ and so correspond to a particle spectral index of about $p\sim
2.2$. Using Fig.~\ref{fr2} it can be estimated that a spectrum with an index
of $2.2$ could already be explored by the three year sensitivity of {\sc IceCube}.
\subsubsection{Flat Spectrum Radio Quasars}
Using a similar correlation between radio and neutrino flux as for FR-II
galaxies, the contribution from FSRQs to the diffuse neutrino flux can be
calculated to
\begin{equation}
E^2\,\frac{dN}{dE}=1.70\cdot\tau_{eff} 10^{-9}\diffunits\,.
\end{equation}
The flux lies well beyond the current diffuse limit and even below the
sensitivity of {\sc IceCube} for an optical depth around $\tau_{eff}=1$. Equivalent
as for the case of FR-II radio galaxies, the upper limit to the optical depth in FSRQs can
be determined to be
\begin{equation}
\tau_{eff}<44
\end{equation}
by using the diffuse {\sc AMANDA} limit. 

The effective optical depth is quite
uncertain in the case of Flat Spectrum Radio Quasars. Just as an example, a
sub-class of FSRQs are GPS, which are compact AGN where it is assumed
that the jet runs into dense matter, yielding a high potential for
proton-photon interaction which results in a high optical depth. However,
other sources like TeV blazars contribute to the source class, being at the
other end with a very low optical depth.
 
With {\sc IceCube}, it will be possible to examine sources with
$\tau_{eff}>2.5$. Consequently, neutrino emission in coincidence with the
radio signal from FSRQs will only be observable in the near future, if the
optical depth of FSRQs is significantly higher than the one for FR-II galaxies.
\section{Examination of source class capabilities \label{class_analysis}}
In the previous sections it has been shown that the stacking method can be
used to increase the diffuse sensitivity of high energy neutrino telescopes to
certain source classes. 

Further catalogs published recently which are not yet part of the stacking search in {\sc AMANDA}
will be discussed in this section. The references for the catalogs are given
throughout the text. They will be examined with respect to the
potential neutrino signal and the effectiveness of the method.

Also, a diffuse component from unresolved sources is expected to show
deviations from a purely isotropic flux, since it is expected to follow the
source distribution in the sky. The total signal from a certain
source class to be observed by a neutrino telescope is highly dependent on the
detector's field of view. We will generically examine source classes concerning
the total flux from the northern respectively from the southern
hemisphere. This gives a qualitative examination of the capabilities of {\sc IceCube} and
{\sc KM3NeT} which is planned to observe the southern hemisphere. The local
supercluster for example is observable from the northern hemisphere and so in
some cases, a significant fraction of the total flux in a sample comes from
that hemisphere. It is shown, however, that there are classes with the
dominant contribution in the southern sky. A further constraint is the limitation of the source catalogs themselves: radio data are mainly
given for the northern hemisphere and there are only a few southern identified
sources in the case of very sensitive radio-selected samples. Here, high energy photon
catalogs which are mostly provided by satellite experiments are much more complete in the sense of directionality.

\subsection{Additional source catalogs}
\begin{figure}[h!]
\centering{
\includegraphics[width=10cm]{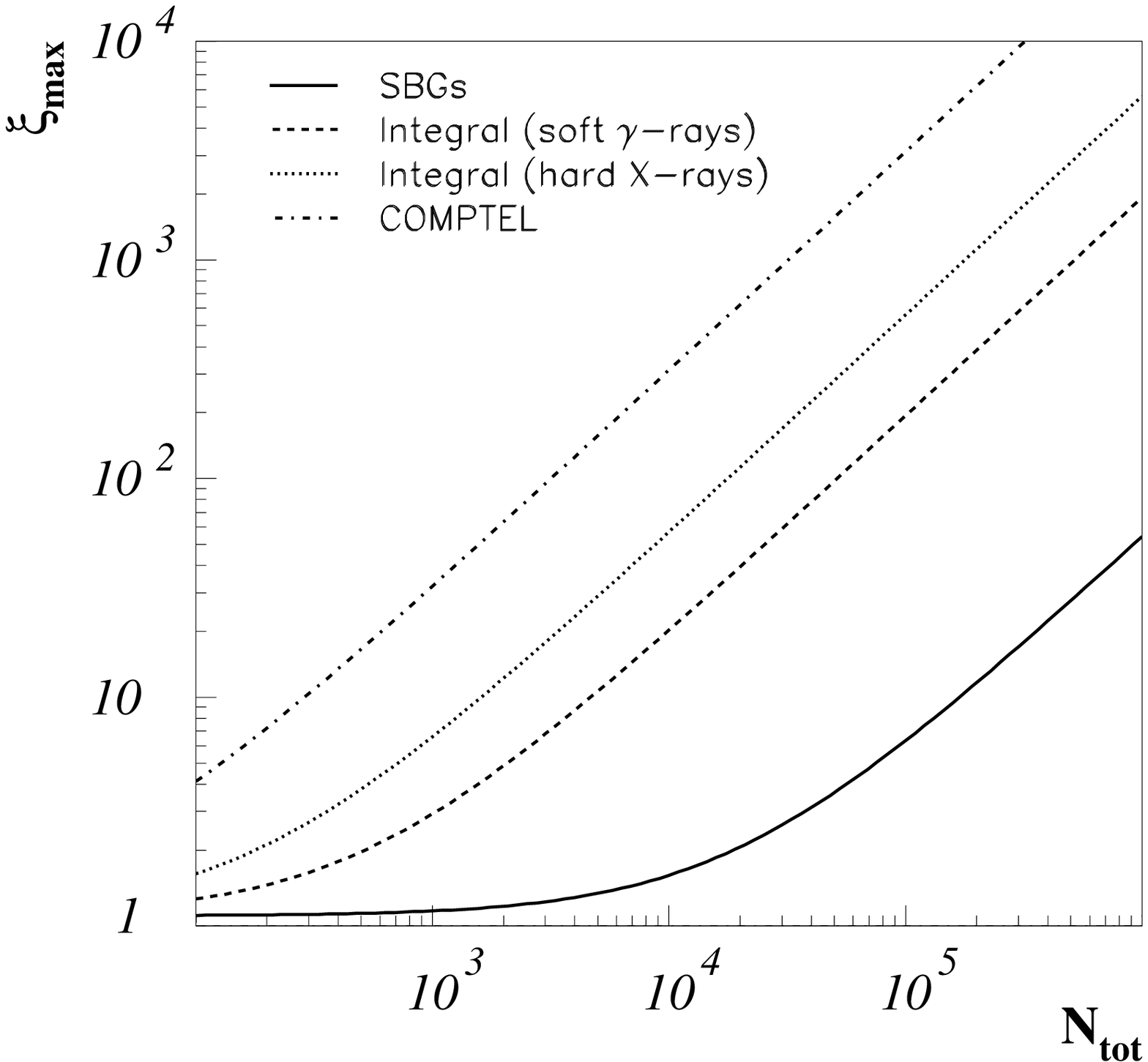}
\hangcaption{Behavior of the maximum diffusive factor with the total number of sources in
  the source class. Only a fraction of the total number of sources is included
in the catalog, dependent on the instrument's sensitivity.}
\label{xi_new_cats}
}
\end{figure}

This subsection gives an overview of four additional source catalogs which yield high
capabilities for neutrino searches with the stacking approach. The point source
sensitivity is likely to be increased in all the cases, diffuse limits can be
derived in two of the cases. A summary of the basic properties of these
catalogs is given in table~\ref{new_cats_basics}. The maximum diffusive factor
for the examined catalogs
can be determined as described in Section~\ref{stack_results},
Equ.~(\ref{ximax}), assuming that the total number of sources in that class
can be determined. This is quite challenging, since a lower luminosity limit
is difficult to determine. Figure~\ref{xi_new_cats} shows the increase of the
diffusive factor with $N_{tot}$ for the whole sky for all four examined
catalogs. Table~\ref{new_cats_prop} reviews the parameters used for the four catalogs.
\begin{table}[h!]
\centering{
\begin{tabular}{l|l|l|l}
Catalog&energy range&Reference&Underlying $\nu$ model\\ \hline\hline
{\sc COMPTEL}&$<100$~MeV&\citep{comptel}&\citep{mannheimjet,stecker_mod}\\
{\sc INTEGRAL}&hard X-ray&\citep{integral_cat}&-\\
{\sc INTEGRAL}&soft $\gamma$-ray&\citep{integral_cat}&-\\
Starburst&FIR&-&\citep{waxbah}\\
\end{tabular}
\caption{Summary of source catalogs interesting to examine with respect to the
neutrino output of the sources.\label{new_cats_basics}
}
}
\end{table} 
\begin{table}[h!]
\centering{
\begin{tabular}{l|l|l|l}
Catalog&$\#(sources)$&$S_{tot}^{cat}$&$S_{weak}$\\
       &south/north&&\\ \hline\hline
{\sc COMPTEL}&11&$0.638\times10^{47}$erg/s/Gpc$^2$&0.0199$\times 10^{47}$erg/s/Gpc$^2$\\
       &5/6&0.326/0.311&0.0384/0.0199\\\hline
{\sc INTEGRAL}&15&$175\times 10^{-11}$~erg/s/cm$^2$&$0.98\times 10^{-11}$~erg/s/cm$^2$\\
(hard X-rays)&10/5&130.08/45.1&0.98/4.34\\ \hline
{\sc INTEGRAL}&42&$441\times 10^{-11}$~erg/s/cm$^2$&$0.85\times 10^{-11}$~erg/s/cm$^2$\\
(soft $\gamma$ rays)&23/19&265/176&1.31/0.85\\\hline
Starburst&199&$17000$~mJy&$0.906$~mJy\\
    &46/153&2260/9480&1.52/0.906\\\hline
\end{tabular}
\caption{Summary of the main parameters in the source catalogs. The total
  number of sources in a catalog is given, $\#(sources)$.
  $S_{tot}^{cat}$ is the integrated flux in the catalog and $S_{weak}$ is
  the measured flux of the weakest source in the sample. Units are given for
  each individual source class. The second row for each class shows the same
  properties for north and south (south/north).\label{new_cats_prop}}
}
\end{table}
\subsubsection{Sources of MeV emission}
Neutrino flux predictions using the diffuse background as measured in hard
X-rays by {\sc COMPTEL} are based on the assumption that the cosmic
diffuse photon flux below 100~MeV is directly correlated to the neutrino
flux. About 10 blazars could be resolved by {\sc COMPTEL} 
underlining the assumption that intrinsically weaker AGN are responsible for
the diffuse signal. No neutrino stacking analysis of the resolved {\sc COMPTEL} blazars has
been done yet. Using the same assumptions as~\cite{mannheimjet}
({\em M95-B}) and \cite{stecker_mod} ({\em StSa05}) would challenge such
an analysis. The derivation of a diffuse limit from a stacking limit would
however be difficult at this stage, since only a negligible fraction of the
sources potentially responsible for a diffuse signal is resolved into sources. Thus, the major fraction of the diffuse signal would
not be included in a stacking analysis and the limit would not apply to diffuse predictions. The
planned launch of satellites such as {\sc GLAST}~\citep{glast} and
{\sc MEGA}\footnote{{\bf M}edium {\bf E}nergy {\bf G}amma-ray {\bf A}stronomy}~\citep{mega}
encourages to pursue the examination of this source
class, since it gives the prospect of increased statistics of resolved sources
in the considered energy range.
\subsubsection{Soft and hard X-rays}
As mentioned before, the limit to a neutrino flux from {\sc ROSAT}-detected sources
would be improved significantly by selecting radio-weak sources instead of
radio strong ones. A significant neutrino flux would not be expected as
discussed in Section~\ref{compare}, but the improvement of the limit would
underline the fact that X-ray emission from the foot of the jet is not
correlated with a neutrino signal.

In hard X-rays, 15 sources have been identified in the {\sc
  INTEGRAL}\footnote{{\bf I}{\bf N}{\bf T}{\bf E}rnational {\bf G}amma-{\bf
    R}ay {\bf A}strophysics {\bf L}aboratory} survey by
JEM-X~\citep{integral_cat} and could be interesting to examine. With the
current analysis, {\sc HEAO-A} results were used, while the more recent measurements
by {\sc INTEGRAL} are bound to be more accurate. Since a direct correlation between hard X-ray sources as detected by Integral and
soft X-ray sources as seen by {\sc ROSAT} does not seem to exist, an
independent analysis would be reasonable. The emission of hard X-rays is
likely to be correlated with TeV photon emission at least via the up-scattering
of synchrotron photons by Inverse Compton scattering. Therefore, under the assumption of a correlated emission of hard X-rays
and TeV photons, an analysis of the hard X-ray signal gives an indication of the flux from high-frequency
peaked BL Lacs.
\subsubsection{Soft $\gamma$-rays}
A catalog of 42 AGN with emission in soft $\gamma$-rays, ($20,100$)~keV has been released by
{\sc INTEGRAL}~\citep{integral_cat}. Here, some of the sources detected by
{\sc OSSE}\footnote{{\bf O}riented {\bf S}cintillation {\bf S}pectrometer {\bf E}xperiment}
in the energy range of ($50,150$)~keV  could be confirmed, others were not seen. We suggest to
use {\sc INTEGRAL} data to define a source class for the stacking of
potential neutrino sources. 
\subsubsection{Starburst Galaxies}
The emission of neutrinos from Starburst Galaxies has been suggested
by~\cite{lw} (L\&W in the following). It is assumed that
\begin{enumerate}
\item Relativistic protons are accelerated along with relativistic electrons,
\item the observed radio emission results from pion-induced electrons. The
  same pions produce neutrinos,
\item protons lose all their energy in $p\,\gamma$ interactions before
  reaching the diffusion time.
\end{enumerate}
It has been pointed out by~\cite{stecker_sbg06}
that L\&W overestimate the fraction of the diffuse far infrared (FIR) flux
coming from Starbursts. While it is $\sim 23\%$ on average, L\&W assume that $100\%$
of the detected signal comes from Starbursts. On the whole, the diffuse flux from
Starbursts is presumable much lower than predicted: L\&W assume that Starbursts are loss
dominated, which means that most primaries interact and do not escape the
source. This enhances the neutrino flux, since basically all protons lose
their energy in $p\,\gamma$ or $p\,p$ interactions and produce
neutrinos. Observations of the spectral radio index of the sources ($\sim
\nu^{-0.8}$) indicate however that Starbursts are in the diffusion limit, indicating
that a negligible fraction of protons interact and only few neutrinos are
produced. This contribution cannot be expected to be observed by {\sc IceCube}.

There is another possibility to expect enhanced neutrino emission
from Starbursts. In the past few years, it could be shown that long GRBs are
typically connected to the explosion of Wolf-Rayet stars into a supernova
Ic. These occur preferably in Star Forming Regions. Thus, a diffuse flux of
GRBs similar to the prediction of~\cite{waxbah} should originate from the direction of these galaxies. There
are two different ways to normalize the diffuse GRB spectrum. One method is to
assume that the observed keV-photon flux is proportional to the neutrino
flux. In that case, the normalization is dependent on the number of observed GRBs
per year. This number is strongly dependent on the instrument and
the number is not very exact. Under the assumption that GRBs accelerate
protons up to the highest energies, $E_p\sim 10^{21}$~eV, the neutrino
spectrum can also be normalized to the flux of ultra high energy cosmic rays
(UHECRs). In this case, the normalization is independent of GRB
observations. It should be kept in mind that the spectral index of the
spectrum still varies from burst to burst - in the model of Waxman\&Bahcall,
an average spectral index has been used. 

It is possible to look for a neutrino signal from GRBs by stacking Star Burst
Galaxies. This method has one advantage over a triggered-GRB search: it is
a systematic search, since independent of GRB
data. A disadvantage is that only nearby events are included, since the
sample of Starbursts only reaches out to redshifts of $z=0.07$. It should
however be possible to use {\sc IRAS}\footnote{{\bf I}nfra{\bf R}ed {\bf
    A}stronomical {\bf S}atellite} data to identify Starburst Galaxies at higher redshift. The search for a GRB
signal from Starbursts should be considered as a systematic search for choked and
undetected GRBs. The
sources can be selected according to their FIR-flux, since this is a measure
of the SN rate in a Starbursts. A higher FIR flux indicates a high star formation rate,
thus more SNe and therefore, also more GRBs.
\subsubsection{General approach to optimize stacking for diffuse
  interpretation}
In order to get diffuse limits from source stacking, it is important to choose
source classes which have information on both resolved sources and diffuse
background. That way, the parameters $\epsilon$ and $\xi$ are easily and
correctly determined. A good example is the {\sc EGRET} catalog, where both diffuse
and resolved emission could be proven to be correlated. Alternatively, the
determination of an upper limit to $\xi$ is possible when working with a
relatively complete catalog of the strongest sources in the sky by using a
generic number of total sources in the source class as it is described in
Section~\ref{stack_results}, Equ.~(\ref{ximax}). 
The diffuse interpretation of source classes like TeV blazars on the other hand is
difficult, since IACTs can barely look for diffuse emission
and the sensitivity of all-sky monitors such as {\sc
  MILAGRO}\footnote{{\bf M}ultiple {\bf I}nstitution {\bf L}os {\bf A}lamos {\bf G}amma {\bf R}ay {\bf O}bservatory} is not high enough yet
to detect extragalactic diffuse emission. Future Projects like {\sc
  HAWC}\footnote{{\bf H}igh {\bf A}ltitude {\bf W}ater {\bf C}herenkov} experiment as a
successor of {\sc MILAGRO}~\citep{hawc,hawc_madison06} and {\sc CTA}\footnote{{\bf C}herenkov {\bf
    T}elescope {\bf A}rray} following in the footsteps of
{\sc H.E.S.S.}\footnote{{\bf H}igh {\bf E}nergy {\bf S}tereoscopic {\bf
    S}ystem} and {\sc MAGIC}\footnote{{\bf M}ajor {\bf A}tmospheric {\bf
    G}amma {\bf I}maging {\bf C}herenkov Telescope}, but aiming at a large field of view~\citep[e.g.]{cta}, will help enhancing
the completeness of the sample.

In the first approach of AGN stacking in order to examine a potential excess
in neutrinos, the number of used sources was determined by optimizing the
significance in the detector. The optimal number of sources was in that case
typically around $\sim 10$. In order to achieve the best stacking diffuse
limit it is interesting to optimize the number of sources by taking into
account $\xi$ as well. It needs to be tested if, by reducing $\xi$ as much as
possible and taking the penalty of a possibly increased point source stacking
limit instead, the diffuse limit can be increased. Also, it needs to be
considered carefully if an optimization to a possible detection or to a limit
is the most reasonable choice.
\subsection{Source class evolution}
For the standard analysis of muon neutrino signatures in high energy neutrino
detectors, the field of view is $2\,\pi$~sr. The two experiments {\sc IceCube} and
{\sc KM3NeT} will be observing the
northern respectively the southern hemisphere in the near future and whole sky
coverage is achieved by the combination of the two experiments. 

Previously, in the stacking analysis of {\sc AMANDA} data, sources with
$\delta<10^{\circ}$ were excluded due to the decreasing sensitivity and high
muon background towards the horizon. Here, we include all sources, for
both northern and southern hemisphere down to $\delta=0^{\circ}$, expecting a
much better sensitivity and better muon rejection near the horizon for both
{\sc IceCube} and {\sc KM3NeT} due to improved directional reconstruction.

In this section, the different source classes discussed above will be examined
with respect to their luminosity evolution. The total sample will be compared
to the contribution from the northern and from the southern sky. For each
sample, two figures will be discussed. (a) represents the differential source
counts, $\log N(>S)$ versus the logarithm of the flux $S$. (b) shows the
integral source evolution with the total flux of $N$ sources, organizing the
sources according to their strength, starting with the brightest one. Open
(black) sources display in both cases the whole sky. Red stars
consider only the southern hemisphere and blue triangles show the northern
hemisphere contribution.
\subsubsection{Optically thick blazars - MeV to GeV emission}
\begin{figure}[h!]
\centering{
\includegraphics[width=\linewidth]{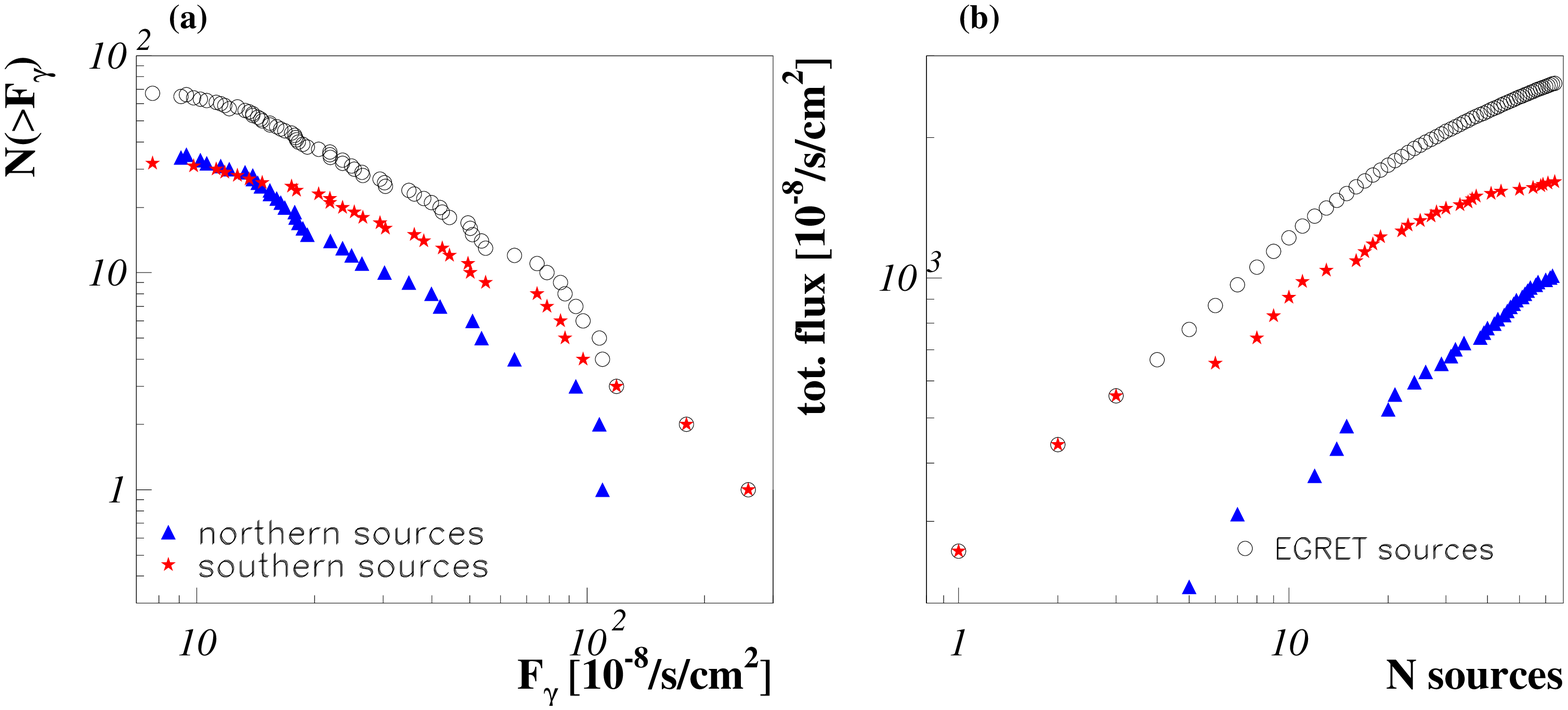}\\[-0.3cm]
\hangcaption[]{(a): Number of sources in the {\sc EGRET} catalog with a flux $>F_{\gamma}$. The
  complete catalog is shown as open circles. The sample has been divided into
  a sub-class of sources at the northern (triangles) and southern (stars)
  hemisphere. (b): Total luminosities of all sources $<N$, starting to sum up
  with the strongest source and successively adding the next luminous source.
\label{egret}\\[-0.8cm]
}
}
\end{figure}

\begin{figure}[h!]
\centering{
\includegraphics[width=\linewidth]{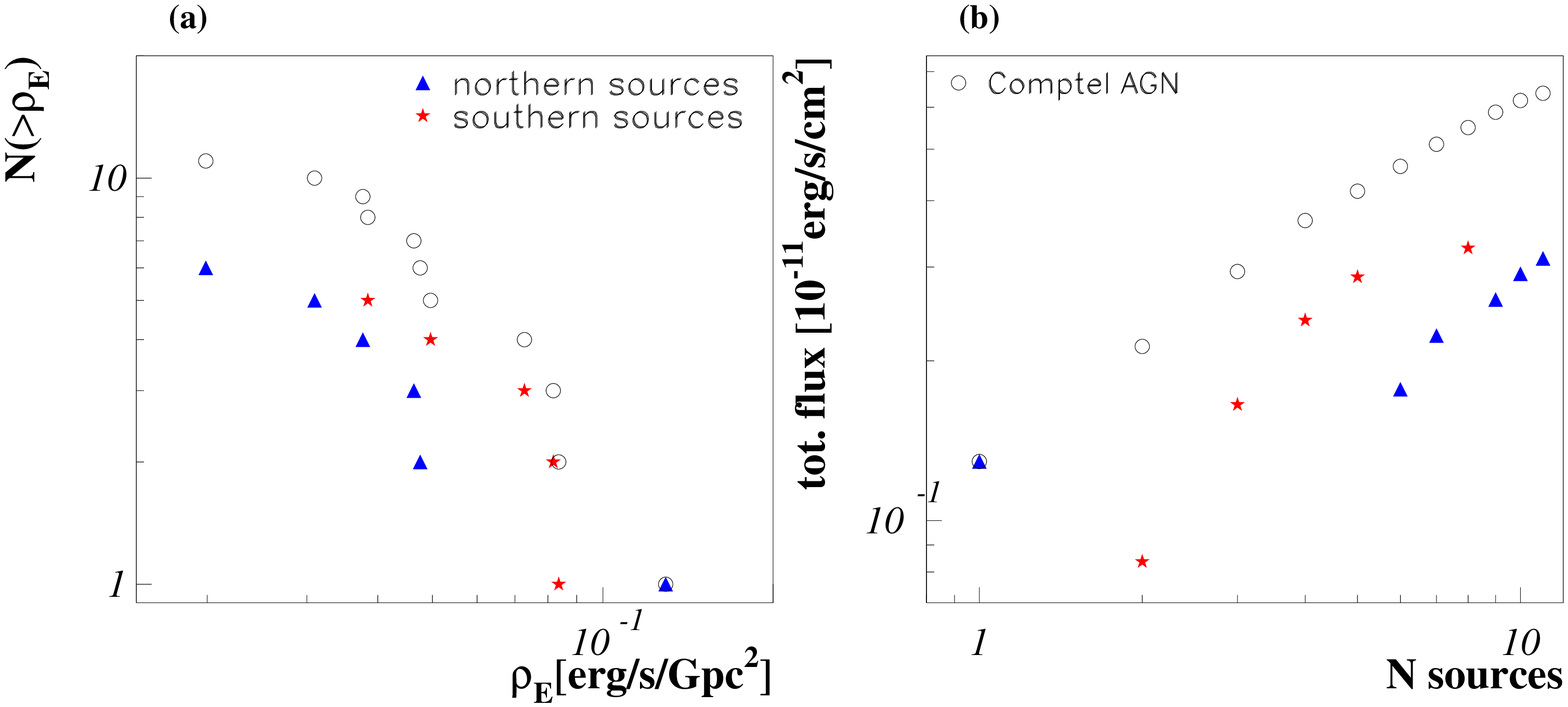}
\hangcaption{(a): Number of sources in the {\sc COMPTEL} catalog with an energy density
 $>\rho_{e}$. 
(b): Total energy density of all sources $<N$, starting to sum up
  with the strongest source and successively adding the next luminous source.
\label{comptel}
}
}
\end{figure}

Figure~\ref{egret} shows the luminosity evolution of {\sc EGRET} sources, for
the whole sky as well as for the northern and southern populations.
The three most luminous sources 
in the {\sc EGRET} sky are in the southern
hemisphere. That is why the main contribution from GeV $\gamma$ rays is
located in the southern hemisphere where the neutrino signal contribution is thus expected
to be much higher. Given the restrictive limit which could already be derived
from {\sc AMANDA} data, this source sample is interesting for {\sc IceCube}. Considering
the source distribution in the sky, the class of GeV blazars is particularly
useful for {\sc KM3NeT} given the high total flux.
\newpage
The {\sc COMPTEL} Catalog of sources with $E<100$~MeV is displayed in
Fig.~\ref{comptel}. In this case, the main contribution lies in the northern
hemisphere. The small number of only 11 identified {\sc COMPTEL} sources make a diffuse
interpretation of a possible limit difficult. Satellites like {\sc MEGA} would make
the investigation even more interesting. The 11 identified sources are however
still useful to investigate with respect to a point source signal. The
maximum diffusive factor lies around $\xi_{\max}\sim 300$ for 10,000 sources and $\xi_{\max}\sim
3000$ for 100,000 sources in the class. Depending on the steepness of the
evolution function, this factor can be significantly smaller, so that
doubling the number of sources could already help to draw a conclusion about
a diffuse limit.\\[-1.cm]
\subsubsection{Soft $\gamma$-rays}
\begin{figure}[h!]
\centering{
\includegraphics[width=\linewidth]{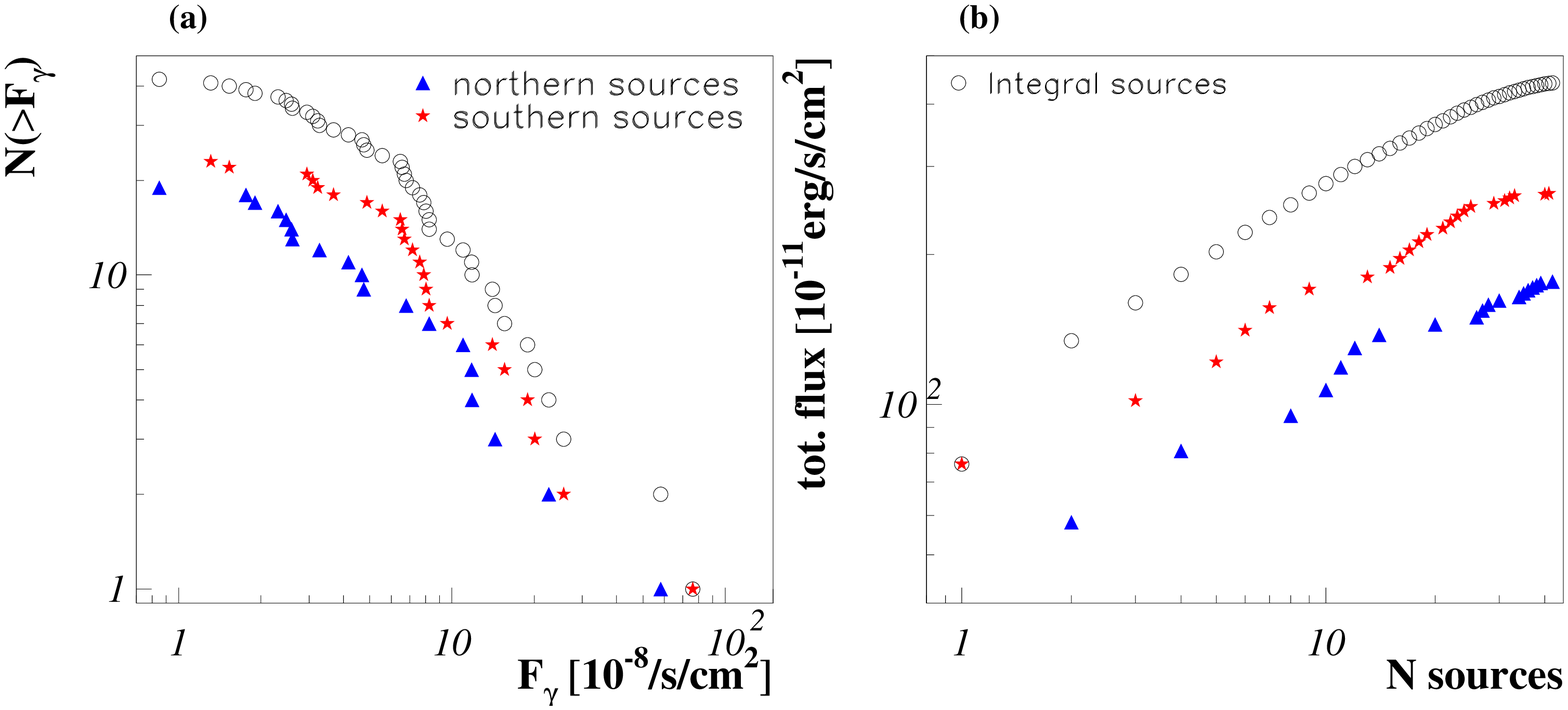}
\hangcaption{(a): $\log N(>S)-\log S$-plot for {\sc INTEGRAL} sources - total catalog
  (circles), only northern sources (triangles) and only southern
  sources(stars). (b): Integral flux versus number of contributing sources.
\label{integral_flux}
}
}
\end{figure}

A catalog of soft $\gamma$-ray-detected AGN is given by
{\sc INTEGRAL}~\citep{integral_cat}. 19 of 42 sources are in the northern sky, among
the 23 southern sources are the
three most luminous ones. The catalog is a good candidate for diffuse
interpretation of the neutrino results, given the relatively high number of
sources. The luminosity evolution is displayed in Fig.~\ref{integral_flux}. It
can be seen that the evolution is still rising and that there is still a
significant fraction of signal missing. This is mirrored in the numerical
value of the maximum diffusive factor which is calculated to $\xi_{\max}\sim
20$ for $N_{tot}=10,000$ sources and $\xi_{\max}\sim 200$ for
$N_{tot}=100,000$ sources. These relatively large numbers increase a potential
limit on the neutrino flux from {\sc INTEGRAL} sources by about one order of
magnitude\footnote{In Section~\ref{stack_results} it is already discussed that
even with a source class of $10^{5}$ sources, the strongest $10^{4}$ AGN make
up the dominant contribution. Thus $\xi_{\max}=20$ is the more realistic value
in this case.}. 

\subsubsection{Soft and hard X-rays}

\begin{figure}[h]
\centering{
\includegraphics[width=\linewidth]{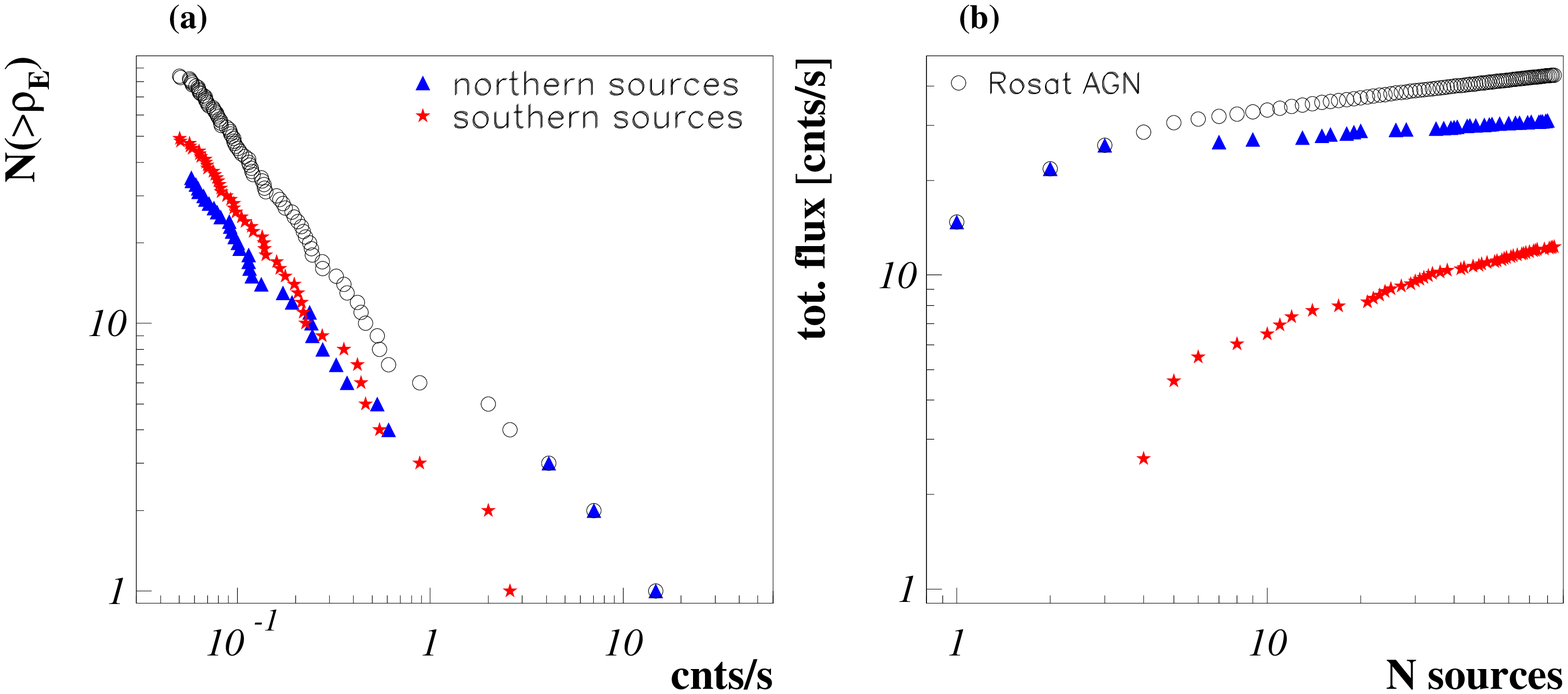}
\hangcaption{(a): Number of sources in the {\sc ROSAT} catalog with an energy density
 $>\rho_{e}$.
(b): Total energy density of all sources $<N$, starting to sum up
  with the strongest source and successively adding the next luminous source.
\label{rosat_cat}
}
}
\end{figure}
Figure~\ref{rosat_cat} shows the luminosity evolution for {\sc ROSAT}-detected
sources. 84 sources are in the sample totally, of which 35 are northern and 49
are
southern sources. The three most luminous sources are in the northern
hemisphere. However, these are objects which have such a large flux, since
they are extremely close to Earth and not because of their high intrinsic
luminosity. This is why these sources were excluded in the stacking analysis
of~\cite{andreas}. Without these sources, the contributions
from both hemispheres are comparable. 

\begin{figure}[h]
\centering{
\includegraphics[width=\linewidth]{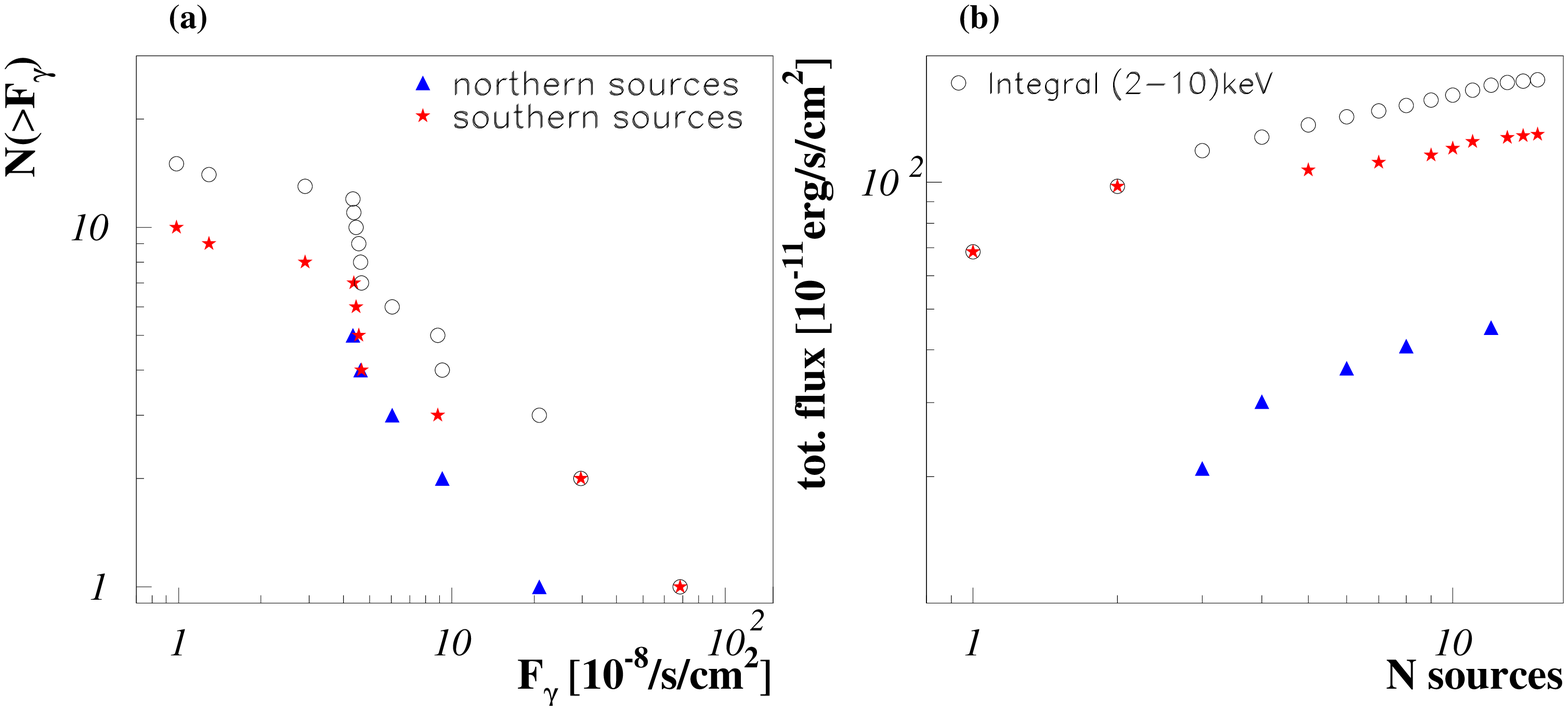}
\hangcaption{(a): $\log N(>S)-\log S$-plot for {\sc INTEGRAL} sources in hard X-rays $(2,10)$~keV - total catalog
  (circles), only northern sources (triangles) and only southern sources
  (stars). (b) total flux of $N$ contributing sources, starting with the
  brightest sources and going to weaker fluxes with higher $N$.
\label{integral_hx}
}
}
\end{figure}
A sample of hard X-ray AGN is available from the {\sc INTEGRAL}
satellite, see Fig.~\ref{integral_hx}. {\sc INTEGRAL} detected 15 sources in the energy band of $2$~keV
to $10$~keV. Among the 10 sources in the southern hemisphere are the two
strongest sources in the sample. There are only 5 northern sources and the
main contribution comes from the southern hemisphere. The {\sc INTEGRAL}
sample can improve the stacking analysis of hard X-ray sources which has been
done with {\sc HEAO-A} information. Only three northern sources were reported from {\sc
  HEAO-A}. For a diffuse interpretation, there are still too few sources in
the sample, though.
\subsubsection{FR-I/FR-II}
\begin{figure}[h]
\centering{
\includegraphics[width=\linewidth]{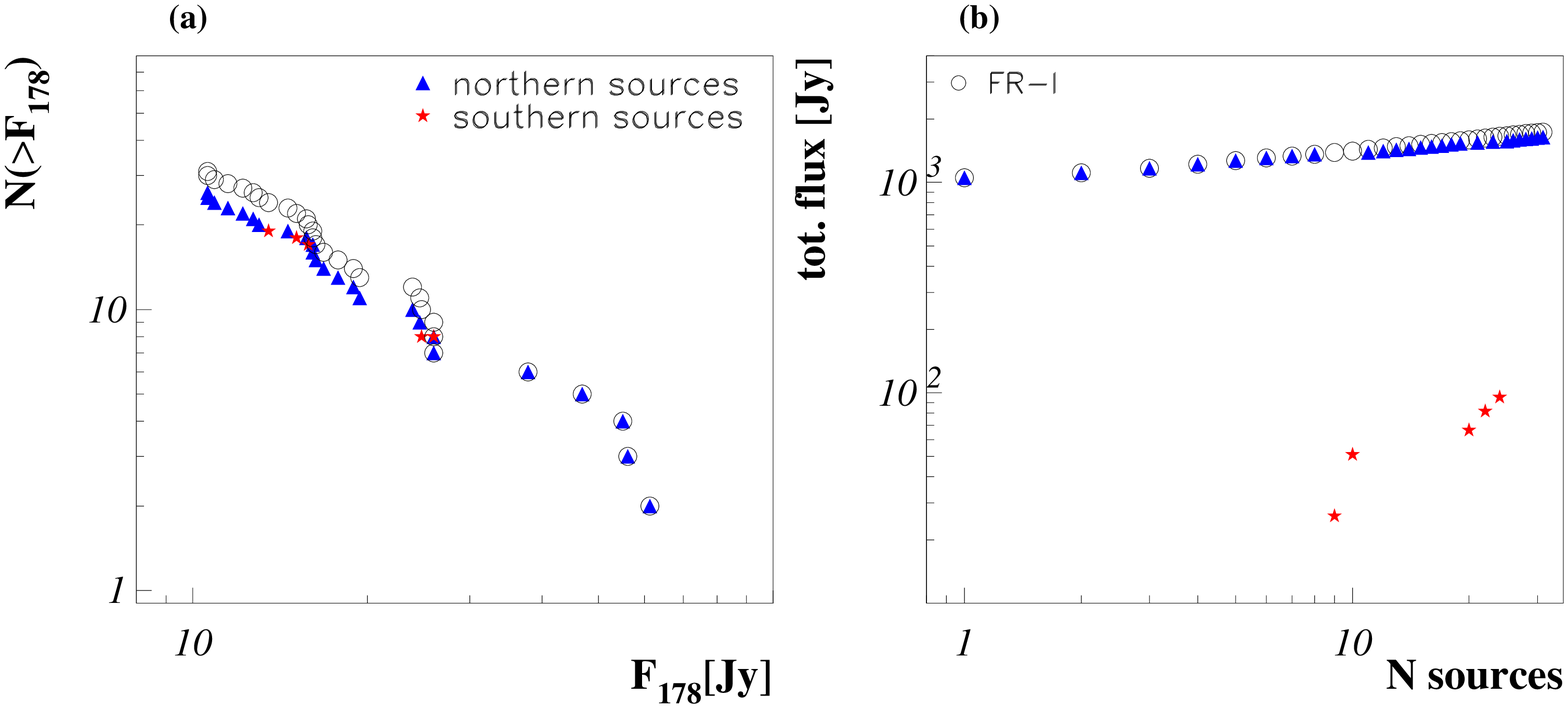}
\hangcaption{(a): Number of sources $>S$ for FR-I galaxies. (b) Total flux versus
  number of sources $N$.
\label{fr_i_cat}
}
}
\end{figure}

\begin{figure}[h]
\centering{
\includegraphics[width=\linewidth]{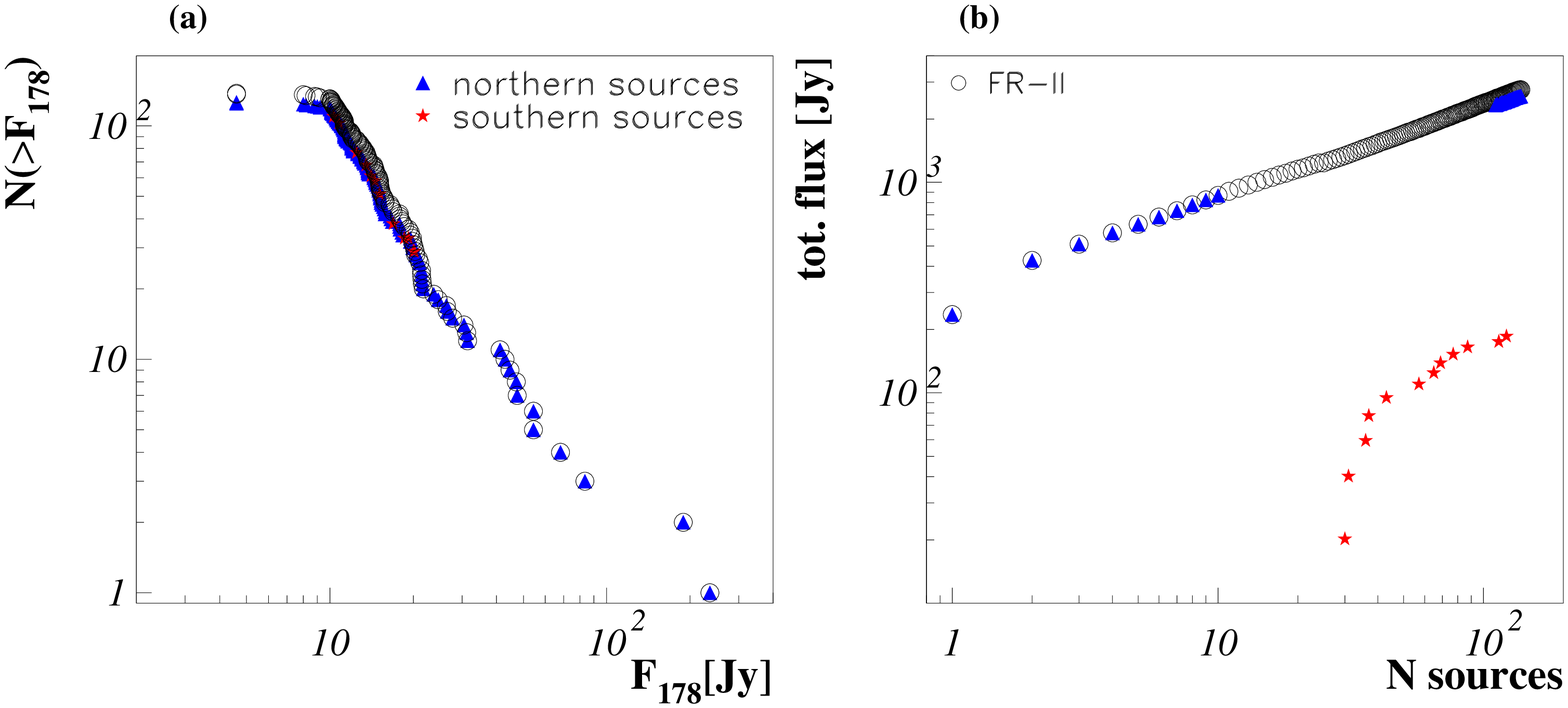}
\hangcaption{(a): Number of sources $>S$ for FR-II galaxies. (b) Total flux versus
  number of sources $N$.
\label{fr_ii_cat}
}
}
\end{figure}

The catalog FR-I and FR-II sources is restricted to values of
$\delta>-10^{\circ}$. Therefore, only few southern sources (7 FR-I and 14 FR-II galaxies) are in the
complete sample which is seen in Fig.~\ref{fr_i_cat} for FR-I galaxies and
in Fig.~\ref{fr_ii_cat} for FR-II galaxies. The flux is totally dominated by
the northern hemisphere.
\subsubsection{CSS/GPS}
\begin{figure}[h]
\centering{
\includegraphics[width=\linewidth]{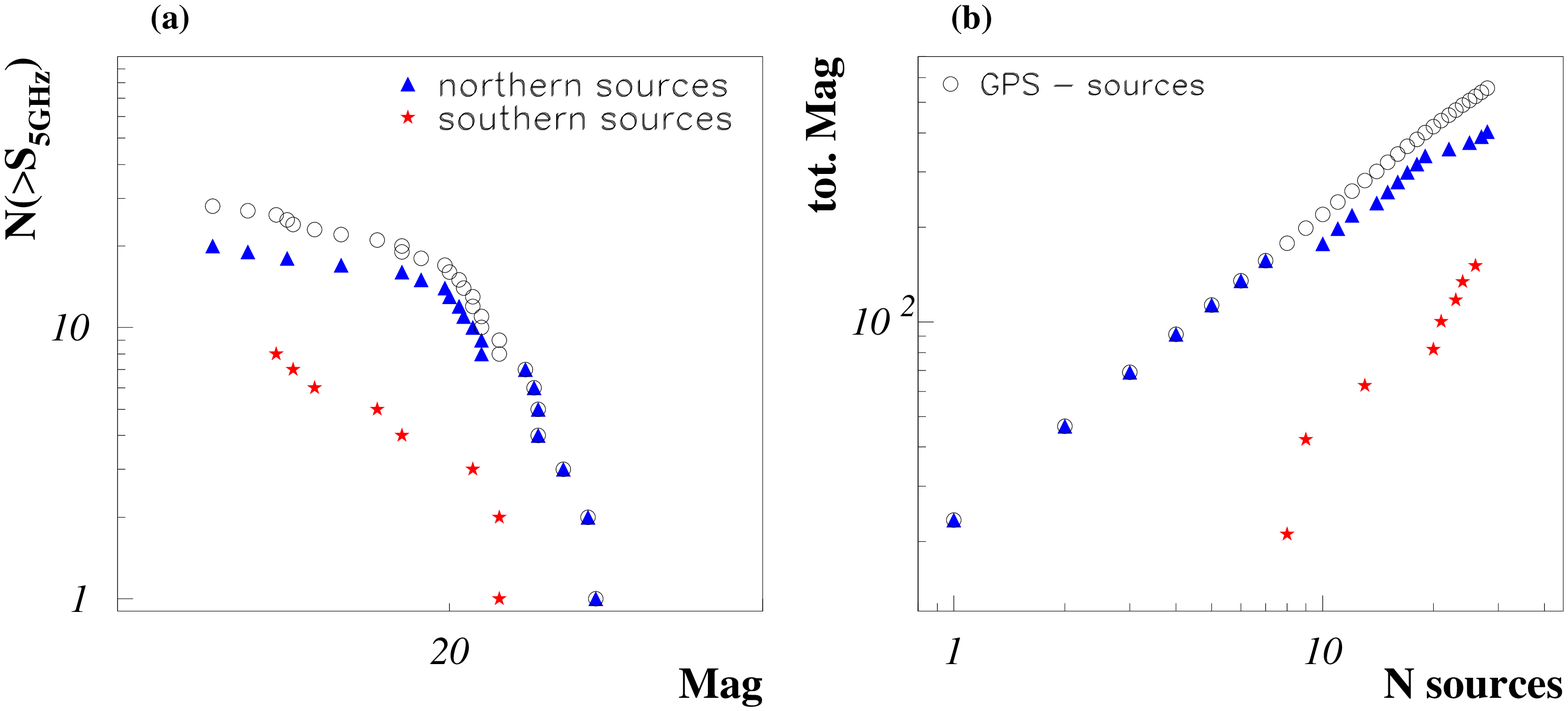}
\hangcaption{Giga-Hertz-Peaked Sources (GPS)\citep{odea}, (a) is the differential
  flux evolution, (b) shows the total flux versus number of sources $N$.
\label{gps_cat}
}
}
\end{figure}

The problem for the catalog of CSS and GPS sources is similar to the situation
of the FR-I and FR-II catalog. CSS have been selected at
$\delta>10^{\circ}$, while GPS include data with
$\delta>-25^{\circ}$. Therefore, there are no southern sources in the case of
CSS. Figure~\ref{gps_cat} shows the GPS sample. 8 southern and 20 northern
sources have been identified with the main contribution from the northern hemisphere.
\subsubsection{QSOs}
In the sample of QSOs as presented by~\cite{sanders89}, there are only 3
sources in the southern hemisphere with a measured flux at the selection
wavelength, $60\,\mu$m. The sample selection of the sample was done for
$\delta>-15^{\circ}$, which only leaves a small window on the southern sky. While this source class is well-suited for the
analysis of the northern hemisphere, there is too little data available in the
southern hemisphere.
\subsubsection{Starburst Galaxies}
The catalog of Starburst Galaxies includes 199 sources.
While the most luminous sources are in the southern, the total flux
is higher in the northern hemisphere, since 153 of 199 sources are located
north. A stacking analysis has good potential for both hemispheres.
\begin{figure}[h]
\centering{
\includegraphics[width=\linewidth]{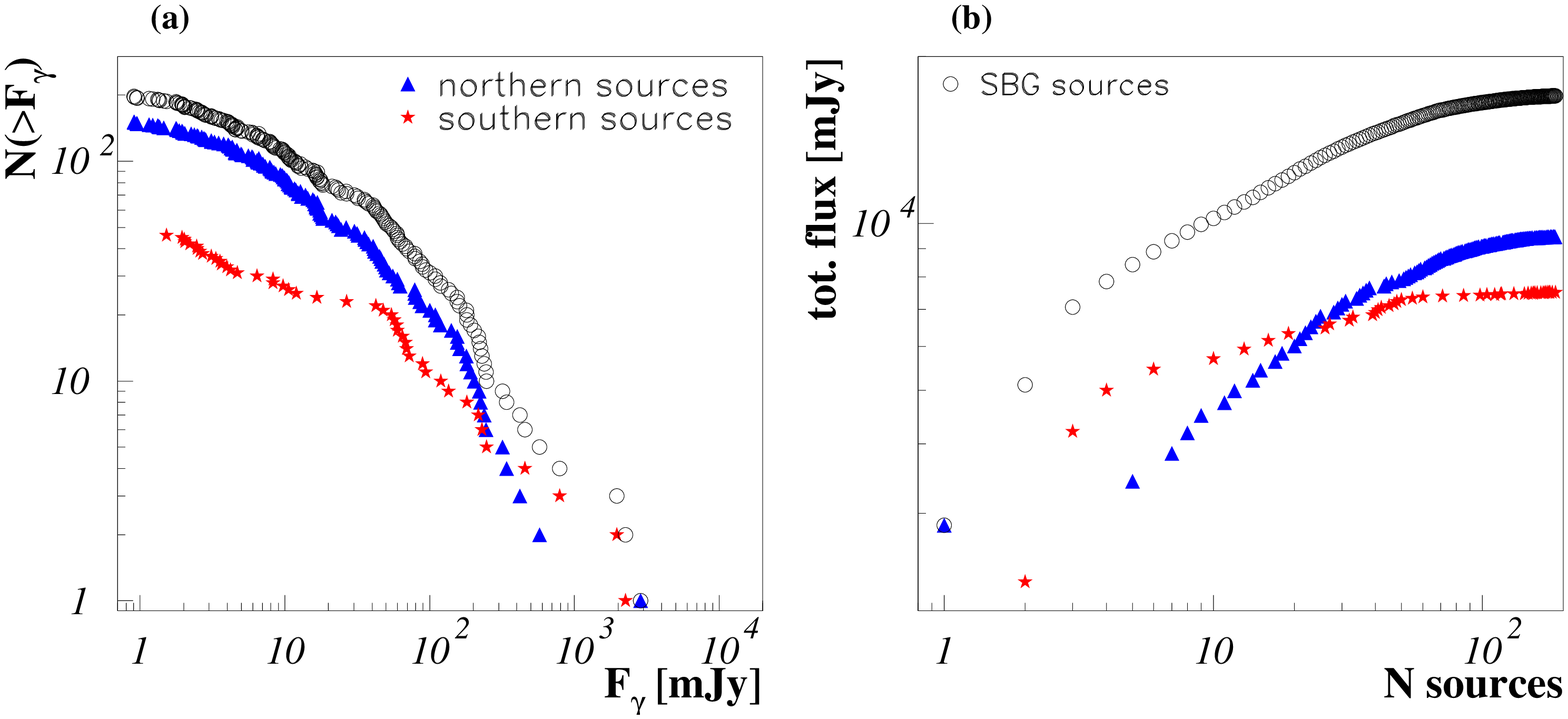}
\hangcaption{(a): Number of sources in the Starburst catalog with a flux $>F_{\gamma}$.
(b): Total flux of all sources $<N$, starting to sum up
  with the strongest source and successively adding the next luminous source.
\label{sbg_cat}
}
}
\end{figure}
\section{Conclusions and Outlook\label{conclusions}}
In this paper, limits to the diffuse flux from seven different AGN source
classes could be derived from point source stacking limits from {\sc
  AMANDA}. There is in many cases an additional constraint on flux models at
the lowest energies: while the general diffuse limit is restricted to the
energy band of $10^{4.2}$~GeV to $10^{6.4}$~GeV, the stacking limits reach down to
$10^3$~GeV. In many cases, the diffusive factor can conservatively only be estimated. It is expected, that the additional diffuse component is much smaller
than the values derived here as upper limits. A reduction of the diffusive
factor would result in the improvement of the stacking diffuse limits.

Neutrino flux predictions normalized to the diffuse {\sc EGRET}
signal above $100$~MeV use the assumption that the diffuse component is
produced by unresolved AGN, see~\citep{mpr,mannheimjet,muecke}. Thus, the {\sc
  EGRET} stacking limit applies to such
calculations. 
The limit falls short of the outline of two of the models,
\citep{mpr,mannheimjet}.
The model of proton-proton
interactions normalized to the {\sc EGRET} diffuse results violates the limit
at low energies around $10^3$~GeV. At the highest energies, the stacking
diffuse limit is more restrictive than the general diffuse limit and reaches
below sensitivities of the maximum contribution from proton-photon interactions.
A detection of a flux from optically thick blazars
should be visible with second generation neutrino telescopes such as {\sc IceCube}. 

The neutrino flux models from X-ray detected AGN and from FR-II galaxies can
be constrained by the general diffuse limit.

In the case of X-ray AGN, the
limit strongly disfavors a hadronic model. It is an order of magnitude below the
flux predictions. According to the strongly restrictive limit, neutrinos are not produced at the foot of
AGN jets in coincidence with X-rays. Thus, the diffuse emission as measured by
{\sc ROSAT} is likely to be due to Inverse Compton radiation.

Analyzing the correlation between the radio and neutrino emission from FR-II galaxies
and blazars, the limit was used to constrain the optical depth of the
sources. For FR-II galaxies, $\tau_{eff}<0.5$ could be derived, while the
upper limit for blazars is given as $\tau_{eff}< 44$. Thus, the detection
potential for such a source class in {\sc IceCube} is very high, since within three
years, {\sc IceCube} is sensitive to sources of optical depth with $\tau_{eff}> 0.024$
(FR-II) resp.~ $\tau_{eff}> 2.1$ (blazars).

It was shown that a general diffuse analysis gives a high discovery potential for hidden TeV
sources: While TeV photons are absorbed at high redshifts, neutrinos propagate
freely. An upper limit to the neutrino flux from photon-resolved TeV blazars
is determined to be
\begin{equation}
{\dl}_{resolved TeV}=1.37\cdot 10^{-9}\,\diffunits\,.
\end{equation}

The considerations above show that an investigation of a neutrino signal can be used
to help determining different intrinsic parameters of the source
type considered. It could be shown that a stacking analysis yields valuable
information on the diffuse contribution from theses sources. The detection
probability is enlarged significantly in such an approach and there is a high
detection potential with {\sc IceCube} and {\sc KM3NeT}. A cascade stacking
analysis by these neutrino detection arrays should be considered in order to increase the
sensitivity to ultra high energy neutrino fluxes as they are described
by~\cite{muecke}.

Apart from the catalogs which have already been used in the stacking analysis
of {\sc AMANDA}, further source classes are investigated here.
The possibility of a stacking analysis of {\sc COMPTEL}
sources is discussed in this paper as a first approach to examine optically
thick sources with photon emission of $E<100$~MeV. Although no stacking diffuse limit can be derived yet
due to the small percentage of resolved sources, future experiments such as
{\sc GLAST} at an energy range of
$(0.02,300)$~GeV~\citep{glast} and {\sc MEGA} at $(0.4,50)$~MeV~\citep{mega} give
hope to resolve many more sources in the MeV range. Another interesting source
class is the catalog of Starburst Galaxies. With the correlation between
starforming regions and long duration gamma ray bursts, an enhanced neutrino
signal from Starbursts can be expected. The prospects of {\sc IceCube} and {\sc
  KM3NeT} are different considering source stacking. Most catalogs show a bias
to one of the hemispheres. Both neutrino detection arrays can be used to
extract complementary information about the different source classes.

The upper limits derived in this paper constraint
several prevailing neutrino flux models. This underlines the necessity of
calculation on the basis of new developments within astroparticle physics and
numerical approaches. A unified model concerning acceleration processes in AGN
would give the opportunity to examine a potential neutrino signal with respect
to different AGN classes within the same framework. The numerical results
could be applied to individual sources and to diffuse flux measurements. The
interaction between the observation of resolved sources and diffuse photon and
proton components of CRs is of high significance with respect as it could be
shown in this paper.
\ack
We would like to thank Francis Halzen, Tanja Kneiske, Anita Reimer, Reinhard Schlickeiser,
Karl Mannheim, Marek Kowalski, Elisa Resconi, Ty de\-Young, Soeb Razzaque and Werner Collmar for inspiring discussions and helpful
comments on this work. Many thanks also to the entire {\sc IceCube}
Collaboration for useful comments and
suggestions.\\
Support for PLB is coming from the AUGER membership and theory grant
05CU~5PD~1/2 via DESY/BMBF. JKB and WR acknowledge the support by the BMBF,
grant 05 CI5PE1/0 and by the DFG, grant RH 35/2-3.

\end{document}